\documentclass{emulateapj}
\usepackage{amsmath}
\usepackage{rotating}
\slugcomment{}

\begin{document}
\title{The Origin of [OII] in Post-Starburst and Red-Sequence Galaxies in High-Redshift Clusters}
\author{Lemaux, B.C., Lubin, L.M., Shapley, A.\altaffilmark{1}, Kocevski, D., Gal, R.R.\altaffilmark{2}, \& Squires, G. K.\altaffilmark{3}}
\affil{Department of Physics, University of California, Davis, 1 Shields Avenue, Davis, CA 95616, USA}
\email{lemaux@physics.ucdavis.edu}
\altaffiltext{2}{Department of Physics and Astronomy, 430 Portola Plaza, University of California, Los Angeles, CA 90095-1547}
\altaffiltext{2}{University of Hawai'i, Institute for Astronomy, 2680 Woodlawn Drive, Honolulu, HI 96822, USA}
\altaffiltext{3}{California Institute of Technology, M/S 220-6, 1200 E. California Blvd., Pasadena, CA 91125, USA}
\begin{abstract}
We present the first results from a near-IR spectroscopic campaign of the Cl1604 supercluster
at $z\sim0.9$ and the cluster RX J1821.6+6827 at $z\sim0.82$ to investigate the nature of 
[OII] $\lambda$3727\AA\ emission in cluster galaxies at high redshift. Of the 401 members in Cl1604 and RX J1821+6827
confirmed using the Keck II/DEIMOS spectrograph, 131 galaxies have detectable [OII] emission
with no other signs of current star formation activity, as well as strong absorption features indicative of
a well-established older stellar population. The combination of these features suggests that the primary source of [OII] emission in 
these galaxies is \emph{not} a result of star formation processes, but rather due to the presence of a Low-Ionization Nuclear Emission-Line 
Region (LINER) or Seyfert component. Using the NIRSPEC spectrograph on the Keck II 10-m telescope, 19 such galaxies were targeted, 
as well as six additional [OII]-emitting cluster members that exhibited signs of ongoing star formation activity. 
Nearly half ($\sim$47\%) of the 19 [OII]-emitting, absorption-line dominated galaxies exhibit [OII] to H$\alpha$ equivalent width (EW) 
ratios higher than unity, the typical observed value for star-forming galaxies, with an EW distribution similar to that observed for LINERs at 
low redshift. A majority ($\sim$68\%) of these 19 galaxies 
are classified as LINER/Seyfert based primarily on the emission-line ratio of [NII] $\lambda$6584\AA\ and H$\alpha$. 
The fraction of LINER/Seyferts increases to $\sim$85\% for red [OII]-emitting, absorption-line dominated galaxies.
The LINER/Seyfert galaxies in our Cl1604 sample exhibit average $L$([OII])/$L$(H$\alpha$) ratios that are significantly higher than that observed in 
populations of star-forming galaxies, suggesting that [OII] is a poor indicator of star formation in a significant fraction of high-redshift cluster members. 
From the prevalence of [OII emitting, absorption-line dominated galaxies in both systems and the fraction of such galaxies that are 
classified as LINER/Seyfert, we estimate that at least $\sim$20\% of galaxies in high-redshift clusters with $M_{\star}>10^{10}-10^{10.5}$ $M_{\odot}$ contain a 
LINER/Seyfert component that can be revealed with line ratios. We also investigate the effect such a population has on the 
global star formation rate of cluster galaxies and the post-starburst fraction, concluding that LINER/Seyferts must be accounted
for if these quantities are to be physically meaningful.

\end{abstract}
\section{Introduction}

At low redshift, the final result of galaxy processing and gas depletion in cluster galaxies is widely observed. Most 
galaxy populations in low-redshift clusters are dominated by bright early-type galaxies, primarily devoid of 
star formation (Dressler et al.\ 1985, 2004; Balogh et al.\ 1997; Hashimoto et al.\ 1998; Lewis et al.\ 2002; G{\'o}mez et al. 2003; 
Pimbblet et al.\ 2006). At higher redshifts ($z\sim$ 0.4-1) where 
this processing has had less time to occur, the fraction of late-type and active or recently star-forming galaxies increases
(Dressler \& Gunn 1988; Couch et al.\ 1994; Dressler et al.\ 1997, 2004, 2009; van Dokkum et al.\ 2000; Lubin et al.\ 2002; Poggianti et al.\ 2006; Oemler
et al.\ 2009). However, the physical processes that are responsible for the quenching of star formation and the transformation of 
disk galaxies to dormant spheroids over the last $\sim$7 Gyr are still not well understood. 

To accurately quantify this evolution, it is essential to use diagnostics that are valid and accessible 
across a broad redshift range. To determine the rate at which a galaxy is forming stars, 
the H$\alpha$ line at 6563\AA\ is typically used, as
it is a relatively dust-independent measure of the star formation rate (SFR) in the last 10 Myr. As H$\alpha$
moves out of the optical window other spectral lines must be used to determine
galaxy SFRs. Many higher redshift surveys (0.6$\leq z \leq$1.4) use instead the [OII] doublet at 3727\AA\ as a proxy for H$\alpha$ 
since it is traditionally associated with nebular regions of current star formation
and is less sensitive to stellar absorption than higher order Balmer lines (e.g., Cooper et al.\ 2006; Vergani et al.\ 2008).

However, a comprehensive study by Yan et al.\ (2006; hereafter Y06) of 55,000 Sloan Digital Sky Survey (SDSS) galaxies at low redshift (0.07$\leq z 
\leq$ 0.1) suggests that [OII] emission is a poor indicator of the SFR in many galaxies. While a large fraction of blue star-forming galaxies 
in the sample have appreciable [OII] emission, approximately 40\% of the red, early-type galaxies also show 
moderate to strong [OII] emission. In 91\% of the latter the [OII] emission likely does not originate from 
normal star formation processes. Rather, the strengths of [NII] $\lambda$6584\AA\ relative to H$\alpha$ and [OIII] $\lambda$5007\AA\ relative 
to H$\beta$ $\lambda$4861\AA\ (i.e., a BPT diagram; Baldwin, Phillips, \& Terlevich 1981) for those [OII]-emitting red-sequence galaxies with all five features 
detected indicate that the line emission is related either to active galactic nuclei (AGNs) or other galactic processes not 
associated with star formation. Most commonly, the source of [OII] emission in these galaxies is related to processes associated with
low-ionization nuclear emission line regions (LINERs). 

This result has significant consequences for galaxy evolution studies. For galaxy populations that are currently forming stars in addition 
to having LINER activity, using [OII] as an SFR indicator results in an overestimate of the global SFR. More importantly, since [OII] probes 
both the star formation and LINER activity, accurately identifying galaxies in a transitory phase
following the truncation of a star formation event (i.e., ``K+A" or ``E+A" galaxies; Dressler \& Gunn 1983; Dressler \& Gunn 1992) 
becomes more difficult when [OII] is used as the sole SFR indicator. Although rare ($\la$2\%)
among bright galaxies both in nearby clusters and in the local and distant field populations (Zabludoff et al.\ 1996; Dessler et al.\ 1999; Goto
et al.\ 2003; Tran et al.\ 2003; Quintero et al.\ 2004; Yan et al.\ 2009), K+A galaxies typically comprise a significant fraction (15\%-25\%) 
of the galaxy populations in distant clusters (Dressler et al.\ 1999, 2004; Tran et al.\ 2003; Oemler et al.\ 2009). 
The correct classification of such galaxies is a vital step in linking the large number of star-forming, disk galaxies 
seen in high-redshift clusters to the quiescent, early-type galaxies observed in their local counterparts (e.g.; Poggianti et al.\ 1999; 
Oemler et al.\ 2009; Wild et al.\ 2009).

The K+A classification is based on two physical properties of galaxies: significant recent star formation and the 
absence of current star formation. The second criterion as applied to high-redshift galaxies ($z\ga0.3$) typically requires that 
the galaxy spectra be essentially devoid of [OII] emission (Dressler \& Gunn 1992; Zabludoff et al.\ 1996; 
Balogh et al.\ 1999; Dressler et al.\ 2004; Oemler et al.\ 2009). 
If this phase of LINER emission is a typical stage of galaxy evolution, excluding all [OII] emitters
from K+A samples severely underestimates the fraction of galaxies that are truly ``post-starburst" or ``post-star-forming". 
In the low redshift sample of Y06 a large fraction ($\sim$80\%) of the K+A population (selected to exclude current star formation
on the basis of the H$\alpha$ line) show appreciable levels of [OII] emission. If cluster galaxies at high redshift
share similar properties, it is necessary to understand where galaxies that exhibit LINER emission lie along the evolutionary 
chain in clusters and what role, if any, LINER emission has in truncating star formation in cluster
galaxies. 

To study the properties of this phenomenon at high redshift we use the extensive spectroscopic database
from the Observations of Redshift Evolution in Large Scale Environments survey (ORELSE; Lubin et al.\ 2009, hereafter L09). 
The ORELSE survey is an ongoing multi-wavelength campaign mapping out the environmental effects on galaxy
evolution in the large scale structures surrounding 20 known clusters at moderate redshift ($0.6 \leq z \leq 1.3$). In 
particular this paper focuses on two structures, the Cl1604 supercluster at $z\approx0.9$ and 
the X-ray-selected cluster RX J1821.6+6827 at $z\approx0.82$. Combining the wealth of previous ORELSE observations 
in these fields with newly obtained Keck II Near-Infrared Echelle Spectrograph (NIRSPEC) spectroscopy of 25 galaxies, 
we investigate the pervasiveness of LINER emission in cluster galaxies at high redshift and the
properties of galaxies whose optical emission lines are dominated by this phenomenon. 

The remainder of the paper is organized as follows: in \S2 we discuss the two high redshift structures targeted
by this survey. \S3 describes the optical spectroscopy and discusses the target selection, 
observation, and reduction of the near-infrared spectroscopy. In \S4 we describe our methods for equivalent 
width (EW), relative flux, and absolute flux measurements, including absolute spectrophotometric calibration and
extinction corrections. In \S5 we present our results and discuss their consequences for  
high redshift galaxy surveys. \S6 presents our conclusions. We adopt a standard concordance $\Lambda$CDM
cosmology with $H_{0}$ = 70 km s$^{-1}$, $\Omega_{\Lambda}$ = 0.7, and $\Omega_{M}$ = 0.3. All EW measurements
are presented in the rest frame and all magnitudes are given in the AB system (Oke \& Gunn 1983; Fukugita et al.\ 1996).
 
\section{Targeted Structures}

To quantify the frequency of LINER emission in cluster galaxies at high redshift, 
we study the galaxy population in the optically-selected Cl1604 supercluster at $z\approx0.9$.
The supercluster is a massive collection of eight or more constituent groups and clusters, 
spanning 13 $h^{-1}$ comoving Mpc in the transverse dimensions and nearly 100 $h^{-1}$ 
comoving Mpc in the radial dimension (see Gal et al.\ 2008 for details, hereafter G08). Additionally, we target galaxies
in the X-ray-selected cluster RX J1821.6+6827 (i.e., NEP5281) at $z\approx0.82$. The properties and data available 
for each structure are discussed individually below.

\subsection{The Cl1604 Supercluster}

The Cl1604 supercluster consists of structures that 
range from rich, virialized clusters dominated by red, early-type galaxies and a hot intracluster medium 
(clusters Cl1604+4304 and Cl1604+4314; hereafter clusters A and B) to sparse chains of galaxies dominated by starbursts  
and luminous AGN (e.g., Kocevski et al.\ 2009a,b). The velocity dispersions of the structures in Cl1604
range from 811$\pm$76 km s$^{-1}$ (cluster B) to 313$\pm$41 (Cl1604+4316, cluster C) (G08). The two most massive clusters 
(clusters A \& B) have well measured bolometric X-ray luminosities ($L_{X,Bol}=15.76\pm1.48$ and $11.64\pm1.49 \times 
10^{43}$ h$^{-1}_{70}$ ergs s$^{-1}$, respectively) and X-ray temperatures ($T_{X}=3.50^{+1.82}_{-1.08}$ and 
$1.64^{+0.65}_{-0.45}$ keV), while the other groups and clusters show no evidence
of a hot intracluster medium ($L_{X,bol}\la 7.4\times 10^{43}$ h$^{-1}_{70}$ ergs s$^{-1}$) (Kocevski et al.\ 2009a).

The imaging data on this structure includes Very Large Arrat (VLA; B-array, 20 cm), \emph{Spitzer} IRAC
(3.6/4.5/5.8/8.0 $\mu$m) and MIPS 24 $\mu$m imaging, archival V band Suprimecam imaging, deep
Palomar 5-m $r\arcmin$ $i\arcmin$ z$\arcmin$ $K_{s}$ imaging, 17 \emph{Hubble Space Telescope} (\emph{HST}) 
Advanced Camera for Surveys (ACS) pointings in \emph{F}606\emph{W} and \emph{F}814\emph{W}, and two deep (50 ks) \emph{Chandra} pointings. The 
reduction of some of the multi-wavelength imaging on the supercluster has been 
discussed in other papers (Kocevski et al.\ 2009a, 2009b; Kocevski et al.\ 2010, in preparation). The analysis of these ancillary data 
as it pertains to the current sample will be discussed in a second paper that
focuses on the multi-wavelength and stellar mass properties of galaxies studied 
in this paper (B.C. Lemaux et al.\ 2010, in preparation). 

\subsection{RX J1821.6+6827}

The cluster RX J1821.6+6827 (hereafter RX J1821) at $z\approx0.82$ was originally observed by \emph{ROSAT} (Tr\"{u}mper 1982) 
in the North Ecliptic Pole Survey (Henry et al.\ 2001; Mullis 2001; Gioia et al.\ 2003). Spectroscopic observations
as part of the ORELSE survey (L09) yielded a velocity dispersion of 926$\pm$77 km s$^{-1}$ from 40 cluster members within
1 h$^{-1}$ Mpc of the cluster center, slightly higher than the most massive cluster in Cl1604. The X-ray temperature and bolometric luminosity of the 
cluster (4.7$^{+1.3}_{-0.7}$ keV and 1.17$^{+0.13}_{-0.18}\times10^{45} 
h^{-2}_{70}$ erg s$^{-1}$, respectively) derived from \emph{XMM-Newton} observations suggest that the cluster is reasonably relaxed, as it lies close to
the $\sigma_{v}-T$ relationship observed in virialized clusters. However, the X-ray morphology is elongated (Gioia et al.\ 2004),
and measurable velocity substructure has
been identified in the spectroscopy (see L09), implying that the cluster is still in the process of formation. Still,
RX J1821 represents a higher mass, higher temperature cluster than those in Cl1604, allowing us to measure the pervasiveness 
of LINER activity in high-redshift clusters at significantly different stages in their dynamical evolution.

The wealth of imaging available for this structure is similar to that for Cl1604, including VLA (B array, 20 cm),
\emph{Spitzer} IRAC (3.6/4.5/5.8/8.0 $\mu$m) and MIPS 24/70 $\mu$m imaging, deep
Palomar 5-m $r\arcmin$ $i\arcmin$ $z\arcmin$ and Kitt Peak 4-m $K_{s}$ imaging, and a single deep (50 ks) \emph{Chandra} pointing.

\section{Observations}

\subsection{Optical Spectroscopy}

\subsubsection{Low-Resolution Imaging Spectrometer}

The original spectroscopic campaign in Cl1604 was conducted with LRIS 
on the Keck 10-m telescopes, obtaining spectra of all galaxies with $R 
< 23$ in the vicinity of clusters Cl1604+4304 and Cl1604+4321. 
Further details on the galaxy selection process, observations, and reduction of these
data are given in the original survey paper of Oke et al. (1998).

Following the original survey, a follow-up LRIS spectroscopic campaign consisting
of six slitmasks was undertaken in the Cl1604 field and is described in detail in Gal \& Lubin (2004). 
Since only one of the LRIS spectra from these observations is used in this study, we only 
briefly give the details of the LRIS observations. The LRIS spectroscopic targets in the
follow-up campaign were observed with the 400 l mm$^{-1}$ grating in multi-object slitmask mode, with a
full-width half-maximum (FWHM) resolution of $\sim$7.8\AA\ and a typical wavelength coverage of 5000 to 9000 \AA. In total 
85 high-quality redshifts were obtained with LRIS with 0.84 $\leq$ $z$ $\leq$ 0.96, the 
adopted redshift range of the Cl1604 supercluster. All galaxies in RX J1821 that are used 
for this study were observed with the DEep-Imaging Multi-Object Spectrograph (DEIMOS); thus, we do not include any additional information
on the original LRIS campaign of Gioia et al.\ (2004) for this system. 

\subsubsection{DEep-Imaging Multi-Object Spectrometer}

The bulk of the redshifts in the Cl1604 field come from observations
of 12 slitmasks with the DEIMOS (Faber et
al. 2003) on the Keck II 10-m telescope between May 2003 and June 2007. The
details of the observations and spectroscopic selection are described in G08. Briefly,
slitmasks were observed with the 1200 l mm$^{-1}$ grating with an FWHM 
resolution of $\sim$1.7\AA\ (68 km s$^{-1}$) and a typical wavelength coverage 
of 6385-9015\AA. The spectroscopic targets for these slits were selected based 
on the likelihood of being a cluster member, determined through a series of color and 
magnitude selections based on data obtained from the Palomar 5-m Large Format Camera (LFC; Simcoe et al.\ 2000). 
The slitmasks were observed with differing 
total integration times depending on weather and seeing conditions. Integration times varied from 7200s to 
14400s in seeing that ranged from 0.52$\arcsec$-1.4$\arcsec$. 

In total, 903 total high-quality (Q $\geq$ 3; 
see G08 for detailed explanations of the quality codes) extragalactic DEIMOS spectra were obtained in the Cl1604
field, with 329 within the adopted redshift range of the supercluster. Combining these
results with the LRIS campaigns, 414 high quality spectra have been obtained for members of the 
Cl1604 supercluster. The spectroscopic survey is not explicitly magnitude limited; however, based on the turnover in 
the number counts of galaxies with high-quality spectra, we estimate that our spectroscopic sample 
is representative of all galaxies brighter than $i\arcmin\sim23$ (\emph{F}814\emph{W}$\sim22.5$). The limiting 
magnitude of the spectral sample is much fainter than the turnover magnitude, probing galaxies down to $i\arcmin=25.2$ (F814W$\sim26$). 

Two DEIMOS slitmasks covering the RX J1821 field were observed in September 2005. The spectroscopic
targets for these slits were selected in a nearly identical way to targets in Cl1604. The selection process differed only in 
the LFC color-magnitude cuts that were introduced to account for the small difference 
in rest-frame bandpasses of our filters at the redshifts of the two structures (see L09 for more details). Both slitmasks 
were observed for 5$\times$1800s under photometric conditions, with typical seeing of 0.6$\arcsec$-0.8$\arcsec$. The spectroscopic setup 
(grating used, blocking filter, slit widths, and central wavelength) was identical to that of the Cl1604 DEIMOS observations. 
 
Of the 20 original cluster members observed by Gioia et al.\ (2004) in RX J1821, 
seven were re-observed with DEIMOS. In total, 189 high quality (Q $\geq$ 3) redshifts were obtained from DEIMOS observations of 
the RX J1821 field, with 73 galaxies lying between 0.805 $\leq z \leq$ 0.83. With the 
12 additional redshifts obtained by Gioia et al.\ (2004) within the adopted redshift range, the total 
spectroscopic database for RX J1821 contains 85 cluster members. Similar to Cl1604, the spectroscopic data in this system 
is also representative for galaxies brighter than $i\arcmin\sim23$, with a limiting magnitude of $i\arcmin=24$. 

The exposure frames for each slitmask in the Cl1604 and RX J1821 fields were combined using the DEEP2 version of the 
\emph{spec2d} package (Davis et al.\ 2003). This package combines the individual science exposures of the slitmask and performs 
wavelength calibration, cosmic ray removal, and sky subtraction on a slit-by-slit basis, generating processed two-dimensional 
and one-dimensional spectra for each slit. Further details of the \emph{spec2d} package and the reduction process are given in Lemaux
et al.\ (2009, hereafter Lem09).

Not surprisingly, the DEIMOS spectral properties of the galaxy population in RX J1821 differ appreciably from those in Cl1604. While 63\% of galaxies 
observed with DEIMOS in the Cl1604 supercluster have detectable [OII] $\lambda$3727\AA\ (hereafter [OII]) emission, a clear sign of 
either AGN or star-forming activity, only 36\% of galaxies observed with DEIMOS in RX J1821 show similar activity, typically at  
lower levels. The galaxy population in Cl1604 also seems to have been more active in the past 1 Gyr relative to RX J1821, as probed by
the average strength of the H$\delta$ absorption line. The spectra of the Cl1604 supercluster members contain Balmer absorption strengths typical of
galaxies with significant star formation in the recent past, whereas the spectra of RX J1821 members are, on average, typical of galaxies with no active 
star formation in the last 1 Gyr.

While RX J1821 is more sparsely sampled than Cl1604, we have sub-sampled the Cl1604 DEIMOS spectroscopic data so that it 
is equivalent to that of RX J1821. A significant difference in the fraction of [OII] emitters and 
the strength of the Balmer absorption features between the two fields is still present. This result suggests 
that the variance in mean spectroscopic properties between the two structures reflects true differences in the galaxy populations. 
While we select galaxies with similar DEIMOS/LRIS spectra in Cl1604 and RX J1821 for near-infrared spectroscopy (see \S3.2.1), 
the environments of these galaxies are significantly different. 
If LINER-type processes in galaxies are only induced by very specific processes (i.e. ram-pressure stripping, harassment, galaxy merging, etc.) 
or are limited to very specific stages in a galaxy's evolution, there should also be clear differences in the observed-frame near-infrared 
emission-line properties of the galaxies in the two structures.

\subsection{Near-Infrared Spectroscopy}

\subsubsection{NIRSPEC Target Selection}

Twenty-five galaxies were targeted for follow-up observations in the two structures with the NIRSPEC spectrograph 
(McLean et al.\ 1998) on the Keck II 10-m telescope. Since this sample consists of a small fraction ($\sim$5\%) of the galaxies 
in Cl1604 and RX J1821, the philosophy adopted in the target selection was to maximize our chances 
at successfully detecting LINER-type galaxies with our observations. 

Our highest priority sample (priority 1) are galaxies with spectra that
would be classified as either quiescent (K) or post-starburst (K+A) based on the bulk of the spectral
features, but which also show low to moderately high levels of [OII] emission [2\AA\ $<$ EW([OII]) $<$ 74 \AA, where a positive 
EW corresponds to a feature in emission, see \S4.1 and Appendix A].
The priority 1 sample consists of 109 galaxies in Cl1604 and 23 galaxies in RX J1821, of which 19
were observed with NIRSPEC (17 in Cl1604 and 2 in RX J1821). While 52\% of the galaxies 
in the Cl1604 priority 1 sample would be classified as post-starburst based on the strength of H$\delta$ alone [i.e., EW(H$\delta$) $<$ 
-5 \AA], we did not impose an H$\delta$ cut 
for priority 1 targets. Priority 1 galaxies that were observed with NIRSPEC were primarily selected 
to maximize the total number of priority 1 galaxies in our sample. Thus, priority 1 galaxies with a close ($r<24\arcsec$) priority 1
companion were favored. The consequences of this choice, as well any possible bias that is introduced as a result, are
discussed later in this section.

All priority 1 galaxies have strong Ca H\&K lines, and a majority have Balmer absorption features, suggesting that star formation has 
been suppressed within at least the last $\sim$ 1 Gyr.
While [OII] emission usually precludes the classification of a galaxy as post-starburst or quiescent 
(e.g.; Balogh et al.\ 1999; Dressler et al.\ 1999; Oemler et al.\ 2009; though not always, see Wild et al.\ 2009 and references therein), 
the spectra of these galaxies strongly suggest a different interpretation. Strong Ca H\&K 
features indicate the presence of a well-established older stellar population and are typically absent following a significant star formation event when
the continuum light is dominated by O and B stars. Similarly, emission from \ion{H}{2} regions can mask Balmer absorption features even in the 
case of relatively minor star formation events (e.g, Taniguchi et al.\ 2000), although the strength of the star formation event needed is somewhat 
sensitive to the star formation history of the galaxy. The presence of these features in priority 1 galaxies suggest that they 
are not undergoing star formation episodes and the [OII] emission does not originate from \ion{H}{2} regions. 

The remaining spectral classes (priorities 2-4) are used to place a second target on our slit. 
Priority 2 (91 galaxies in Cl1604, 5 galaxies in RX J1821) are galaxies that exhibit either [OII] in emission 
with no other strong spectral features or [OII] in emission plus strong Ca H\&K, but with other signs of ongoing star formation 
(i.e., at least one of the higher order Balmer lines was observed in emission). Galaxies with no Ca H\&K and obvious signs of ongoing 
star formation and galaxies that have no features in emission were given the lowest priorities 
(priorities 3 and 4 with 52 and 62 galaxies, respectively, in Cl1604; 2 and 42 galaxies, respectively, in RX J1821). 
Six galaxies of lower priorities were observed with NIRSPEC, four in Cl1604 and two in RX J1821.

The remaining 16 galaxies observed with DEIMOS in Cl1604 could not be classified due to reduction artifacts that 
prevented us from accurately measuring the strength of spectral features. Only one galaxy originally targeted by LRIS was chosen 
as a NIRSPEC target due to the lower spectral resolution and lack of flux calibration in these data. The one LRIS target was chosen for NIRSPEC
because it is the brightest red-sequence galaxy in Cl1604 ($z\arcmin$=19.42, \emph{F}814\emph{W}=20.84), is a strong radio emitter, and has
a priority 1 spectrum. The appreciable [OII] emission [EW([OII])=6.3\AA] present in the spectrum of this galaxy was a sufficient 
mystery for us to warrant targeting.

Initially, the selection process for NIRSPEC targets only involved the use of the DEIMOS spectral data.
Following these initial observations, we included broadband color cuts, which was used to differentiate
between galaxies on the observed ACS red sequence and those blueward of it. As nearly member galaxies in
RX J1821 lie on the observed LFC red sequence (see \S5.3), we did not apply any broadband color selection to
the potential NIRSPEC targets in this field. These color criteria were imposed in Cl1604 in order to favor
[OII]-emitting red-sequence galaxies, which were preferred as targets as a large fraction (91\%) of such galaxies at low-redshift exhibit [OII] emission
that is inconsistent or likely inconsistent with normal star-forming processes (Y06). If galaxies at higher redshift exhibit similar trends, selecting
red-sequence galaxies greatly improves our chances of successfully observing galaxies that contain a LINER component.

\begin{figure}
\plotone{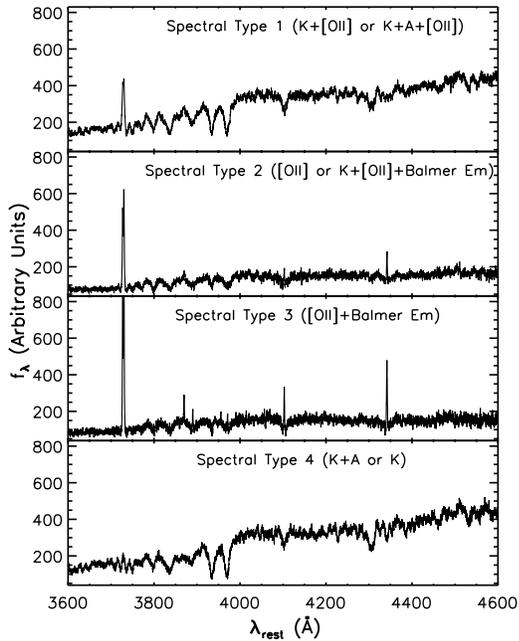}
\caption{DEIMOS spectral co-additions of potential NIRSPEC targets in the Cl1604 supercluster members. The co-additions of four 
priority classes selected primarily by the absence or presence of certain spectral features (see \S3.2.1) are shown.
Each co-addition is a luminosity inverse variance weighted mean. Priority 1 galaxies (top panel, 108 galaxies) exhibit strong
Ca H\&K and Balmer absorption features, as well as a moderately strong 4000\AA\ break, indicative of a dominate 
older stellar population and recently truncated star formation.
However, strong [OII] emission is also present suggesting that these systems may be still forming stars 
or, alternatively, have some contribution from a LINER or Seyfert. Priority 1 galaxies constituted 76\% (19/25) of the 
galaxies targeted with NIRSPEC. Priority 2, 3, and 4 populations, from which the six other targets were drawn, have
co-additions performed with 91, 52, and 62 galaxies, respectively.}
\label{fig:prioritycoadds}
\end{figure}

To demonstrate the differences between the various priority classes used for NIRSPEC observations, we use the full Cl1604 DEIMOS 
spectroscopy. Figure \ref{fig:prioritycoadds} shows the spectral  ``co-addition" of 
the galaxies that comprise each priority class. The co-additions 
were performed in a manner nearly identical to that of Lem09. In order to compensate for the effects of slit-loss, 
the flux in each DEIMOS spectrum was normalized to unity and re-weighted by the absolute 
$i\arcmin$ magnitude of the galaxy. The differences between the composite properties 
of the priority classes can be seen clearly in the spectra.

\begin{figure}
\epsscale{1.25}
\plotone{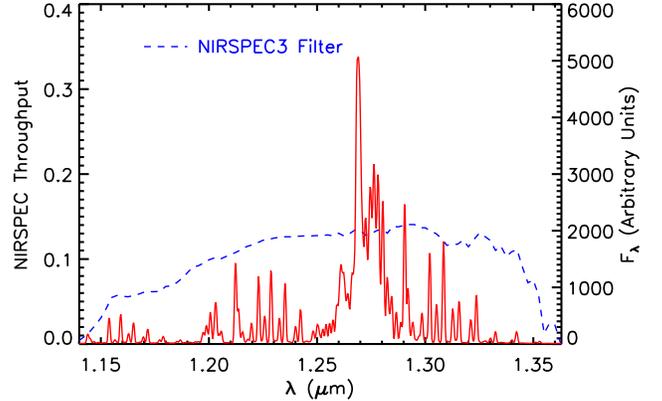}
\caption{Throughput curve of the NIRSPEC3 (J-band) filter against the backdrop
of the near-IR night sky airglow lines. The relative intensity of the OH sky lines has been scaled for 
clarity. NIRSPEC throughput includes light lost from a point source due to the slit as well as losses 
associated with the telescope. All NIRSPEC targets have $z\leq0.93$, which allowed observed-frame 
H$\alpha$ and [NII] $\lambda$6584\AA\ to avoid the strong OH features at 
$\lambda\approx1.27\mu$m.}
\label{fig:Jbandonsky}
\end{figure}

We also required that galaxies have $z\leq0.93$,
so that any observed H$\alpha$ emission is blueward of the strong OH airglow lines at
$\lambda\approx1.27\mu m$ (see Figure \ref{fig:Jbandonsky}). Additionally, we required any potential target to have a nearby
supercluster member that could also be placed on the slit during the observations. This constraint requires that our targets 
have at least one other spectroscopically confirmed supercluster member within 24$\arcsec$. Because the targeted
pairs are kinematically related at a rate similar to all galaxies with similar separations, this
criterion adds no additional bias to our sample as we are sampling galaxies with clustering properties
representative of the majority of supercluster members.

\begin{figure}
\epsscale{1.2}
\plotone{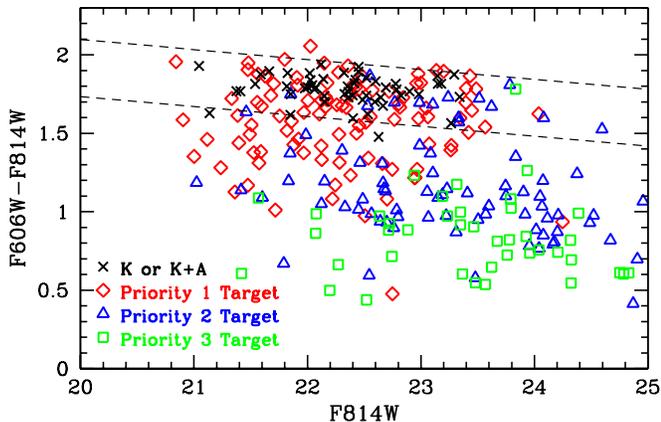}
\caption{Color-magnitude diagram of the 279 DEIMOS confirmed Cl1604 members observed with ACS that were classifiable. Also included is the one LRIS confirmed member 
that was targeted by NIRSPEC. Galaxies with no ongoing star formation (priority 4) are confined almost exclusively to the Cl1604 red sequence. Galaxies that are likely 
undergoing moderate (priority 2) or high (priority 3) levels of star formation activity are primarily found in the faint region of the blue cloud,
though dusty starburst galaxies located on or near the red sequence are among the significant exceptions.
Priority 1 galaxies, which represent the bulk of our NIRSPEC targets, cover a large dynamic range in color-magnitude
space, comprising both the most luminous and least luminous red sequence galaxies as well as a large portion
of bright galaxies with bluer colors.}
\label{fig:priorityCMD}
\end{figure}

Figure \ref{fig:priorityCMD} shows the color-magnitude diagram (CMD) of all four NIRSPEC priority classes 
in the Cl1604 supercluster that have DEIMOS spectra and the one targeted LRIS object covered by our ACS pointings.
The galaxies comprising the three lowest priorities each occupy different, fairly well-defined regions in the CMD.
The lowest priority targets (priority 4, labeled "K or K+A") almost exclusively lie on the red sequence
(see \S5.3 for a detailed discussion on how the red sequence is defined). Priority 3 targets primarily occupy the
bluest part of the ``blue cloud" region, and priority 2 targets primarily
occupy the redder boundaries of the blue cloud. Our priority 1 targets are much more expansive, encompassing both the brightest
and dimmest galaxies on the red sequence and the bulk of the bright galaxies blueward of the red sequence. The
large extent of priority 1 galaxies in color-magnitude space suggests that, even though we are selecting galaxies with similar spectroscopic properties,
we may be sampling galaxy populations at different stages in their evolution. Tables \ref{tab:obs1} and \ref{tab:obs2} list 
the details of each galaxy observed with NIRSPEC as well as their priority classes.

\subsubsection{NIRSPEC Observations}

In total, 25 galaxies were observed with NIRSPEC on the dates of 2007 June 4 and 2008 May 21 UTC, 21 in Cl1604 and 4 in 
RX J1821. The observations consisted of 19 priority 1 galaxies (17 in Cl1604 and 2 in RX J1821), hereafter
referred to as our ``main sample" and six priority 2-3 galaxies (4 in Cl1604 and 2 in RX J1821), hereafter 
referred to as our ``filler sample". Observations for both nights were taken 
in low-resolution mode with slit widths of 0.76{\arcsec}, resulting in a pixel scale of 
3 \AA\ pixel$^{-1}$ and an FWHM resolution of $\sim$8\AA. The observations 
were taken through the NIRSPEC-3 filter (similar to \emph{J} band; see Figure \ref{fig:Jbandonsky}), 
with a typical wavelength coverage of 2900 \AA\ and central wavelength of 1.273 $\mu$m. 
Conditions on both nights were photometric and seeing ranged from
0.3$\arcsec$-0.4$\arcsec$ on the first night and 0.4$\arcsec$-0.8$\arcsec$ on the
second night.

The NIRSPEC-3 filter was chosen to maximize the sensitivity of the
instrument at 1.24 $\mu$m, roughly the wavelength of
H$\alpha$ and [NII] at the redshift of the Cl1604 supercluster.
While our spectral coverage included the  
[OI] $\lambda$6300\AA\ feature (hereafter [OI]), our integration times were only long
enough to significantly detect [OI] in galaxies with extremely hard ionizing spectra.
Our average detection significance of $\sim$7$\sigma$ in \emph{F}(H$\alpha$) implies
a $\sim$1.75$\sigma$ significance of the [OI] feature for typical LINER spectra, too weak to use as a
meaningful diagnostic.

H$\alpha$ and [NII] were chosen to discriminate between star formation and LINER or Seyfert emission. 
This distinction would not be possible if we had instead chosen to observe H$\beta$ $\lambda$4861\AA\ 
and [OIII] $\lambda$5007\AA\ (hereafter [OIII]). Unfortunately, H$\alpha$ and [NII] provide little power to
discriminate between LINERs and Seyferts, which is typically done (when using spectral techniques) by the
ratio of H$\beta$ to [OIII] or the ratio of [OII] to [OIII]. While we attempt to separate 
the two populations using spectral diagnostics in this paper (see \S5.2), our multi-wavelength data 
will be useful in this regard and will be used to fully characterize the nature of such processes in a 
future paper.

Spectral setups typically consisted of two targets observed simultaneously. 
For each setup, we acquired targets by blind offsets from bright ($r\arcmin \sim 17.5$) stars. 
The observation of each setup consisted of staggering 900 s exposures between nods on the sky 
of 1.4$\arcsec$ - 2.5$\arcsec$ along the 42$\arcsec$ slit. A different number of exposures 
were taken for each setup to achieve a similar emission line signal-to-noise ratio (S/N), with the total  
integration times varying between 1800 and 3600 s. Preference was given to setups in the Cl1604 
supercluster, with galaxies in RX J1821 observed only when the Cl1604 field was 
unavailable. Two standard stars drawn from the UKIRT list of bright 
standards\footnote{http://www.jach.hawaii.edu/UKIRT/astronomy/calib/phot\_cal/
ukirt\_stds.html} were observed on each night, HD105601 (A2) at evening twilight and 
HD203856 (A0) at morning twilight. 

\subsubsection{NIRSPEC data reduction} 

The NIRSPEC data were reduced using a combination of Image Reduction and Analysis Facility (IRAF; Tody 1993)
scripts and a publicly available semi-automated Interactive Data Language (IDL) pipeline 
(G. Becker, private communication, 2009). The IRAF scripts (described in detail in Erb et al.\ 
2003) were used primarily to create stacked two-dimensional spectra for use as a visual guide throughout the 
reduction and analysis of the data. All results in this paper are presented from data reduced with the IDL 
pipeline. 

The pipeline initially determines the position of an object on the slit as a function of position on the
1024$\times$1024 NIRSPEC detector. A ``slit-map", made by interpolating between observations of a 
standard star, is used to correct for variations of the spatial position of the science object along the 42$\arcsec$
slit. During this process an initial wavelength solution is also generated using the sky lines in the 
observations of the standard stars. 

Each science frame was processed by differencing the flat-fielded, 
dark-subtracted science frame with a reference science frame. A second wavelength solution was performed by manually 
identifying several cleanly separated bright J-band airglow lines in the science frames. The resultant 
product is a cosmic ray cleaned, wavelength calibrated, dark-subtracted, flat-fielded two-dimensional
spectrum. A one-dimensional spectrum is generated by collapsing the dispersion axis 
and fitting an optimal Gaussian to observed peaks in the brightness distribution. At each wavelength position, the 
flux values for the one-dimensional spectrum are calculated by summing the pixels along the spatial axis, using the best-fit Gaussian 
parameters to optimally weight (Horne 1986) the relative contribution from each pixel. In objects 
where the continuum had low S/N, but emission features were detected, the dispersion axis was collapsed only over the wavelength 
range containing the emission lines. For each science object, these one-dimensional spectra were 
combined into a single spectrum using an inverse variance weighted mean that preserved the overall 
flux calibration. Figures \ref{fig:mosaic1}-\ref{fig:mosaic5} show cutouts of the 
reduced NIRSPEC and DEIMOS/LRIS spectra of all 25 targets, as well as their associated 
ACS \emph{F}814\emph{W} or LFC $i\arcmin$ postage stamps.

\section{Spectral Line Measurements}

\subsection{Equivalent Width and Emission Line Flux Ratios}

In each processed one-dimensional DEIMOS and NIRSPEC spectrum, we measure the EW of the [OII] and H$\alpha$ nebular 
emission features. These EW measurements are 
useful because there exist well measured correlations between the EW of the H$\alpha$ and [OII] lines in
LINER-type and star-forming galaxies at low redshift, which generally are observed to be tighter than correlations observed in
line luminosities (Y06).

In many galaxies the properties of the dust affecting the stellar continuum are appreciably different from the 
nebular dust properties (see Calzetti et al.\ 1994; Calzetti 2001 and references therein). 
This differential extinction causes EWs to be slightly affected by dust abundance, decreasing EW(H$\alpha)$ by a factor of 1.25 and EW([OII]) by 1.36
using $E(B-V)$=0.3 and the Calzetti et al.\ (2000) reddening law. While the absolute numbers may be underestimated, we ignore this effect
as any correlations that exist between the EW of the two features should remain essentially invariant with respect to dust properties. 

The EW was measured in the rest-frame using two different techniques: bandpass measurements and line-fitting techniques. 
Bandpass measurements were performed on all DEIMOS and NIRSPEC spectra by defining three bandpasses in the vicinity of the 
spectral feature adopted from Fisher et al.\ (1998) for [OII] and Y06 for the H$\alpha$ and [NII] features. Line-fitting
was performed on all spectra where emission lines were detected at a significance of greater than 3$\sigma$. For all measurements,
spectra were fit by a double Gaussian model at fixed wavelength separation with a linear continuum. In NIRSPEC spectra we tested 
the effect of including a third Gaussian (at fixed separation) to account for the weaker [NII] $\lambda$6548\AA\ feature. 
In all cases, the effect on EW(H$\alpha$) was extremely small ($\sim$5\%) and consistent within the errors.
For each spectrum, the EW measurement was chosen from the better of the two methods, typically depending on the feature S/N. 
The criteria for this choice as well as further details of the two methods are given in Appendix A. Note that, given our definition
of EW in Appendix A (see Equation \ref{eqn:EW}), the convention adopted in this paper is for positive EWs to correspond to 
features observed in emission and negative EWs to those observed in absorption. 
 
While all [OII] EWs were measured in emission at significances much higher than 3$\sigma$, a subset ($\sim20$\%) of our galaxies had 
H$\alpha$ EWs that were measured in emission at significances less than 3$\sigma$. When analyzing line fluxes such galaxies are typically
excluded (e.g, Tremonti et al.\ 2004; Kewley et al.\ 2006; Y06).
However, our sample is comprised of galaxies that have [OII] in emission at very high significance. Our NIRSPEC observations  
are aimed at finding galaxies with low levels of H$\alpha$ emission relative to their [OII] emission.  Since galaxies with EWs measured at
low significance are, by definition, weak H$\alpha$ emitters (or absorbers), we include the H$\alpha$ EW measurements 
of all galaxies in our analysis regardless of the significance. Table \ref{tab:spec} lists the H$\alpha$ and [OII] EW measurements 
of all galaxies in our sample. 

Emission line fluxes of the [OII], H$\alpha$ and [NII] features were calculated using the same two methods as those used 
for EW measurements. For all galaxies in our sample
we used the same method to determine the line flux as was used for that galaxy's EW measurement. The bandpasses chosen for all features
were identical to those used for measuring EWs. While we include EW measurements detected in emission at a significance less 
than 3$\sigma$ in our analyses, we typically do not 
include low significance measurements of line fluxes. The two exceptions are the [NII] measurement of galaxies 11 and 19, for which 
the line is detected at a significance of approximately 2$\sigma$. In these cases, the [NII] line can be clearly seen
in the two-dimensional spectrum, but the formal error (due to bright airglow lines in the vicinity of the detection) 
places the measurement at a significance of less than 3$\sigma$. The classification of these galaxies does not depend on this choice.
For all other line fluxes that were detected at a significance less than 3$\sigma$ 
(5/25 H$\alpha$ lines; 3/25 [NII] lines), 3$\sigma$ line fluxes were adopted as the formal upper bound. 
These upper limits allowed us to classify the one galaxy for which the H$\alpha$ feature was detected at a significance greater 
than 3$\sigma$ but for which the [NII] line was not (galaxy 7), and the three galaxies (13, 15, and 21) for 
which the [NII] feature was detected at high significance but H$\alpha$ was not. The two galaxies 
(16 and 17) where neither line was detected at a significance greater than 3$\sigma$ are excluded from emission 
line ratio analyses and are discussed further in \S5.2 and \S5.3. Table \ref{tab:spec} lists the H$\alpha$ and [NII]
line fluxes of all galaxies in our sample. 

\subsection{Absolute Flux Measurements}

\subsubsection{DEIMOS Flux Calibration}

Absolute flux calibration of the Cl1604 DEIMOS data was obtained in a manner nearly identical
to that of Lem09. Absolute flux calibration
of the RX J1821 DEIMOS data was not performed as the photometry is 
less accurate (L09), and no \emph{HST} ACS data exist in the field. The four 
NIRSPEC targets in this field are, therefore, excluded from analyses involving 
absolute flux measurements.

The DEIMOS spectrum of each Cl1604 member was response corrected using the generalized DEIMOS 
response function\footnote{http://www.ucolick.org/$\sim$ripisc/results.html} and checked for 
accuracy and precision using the methods of Lem09. 
An average slit throughput of $\omega_{\rm{slit}}=0.37$ was adopted for all DEIMOS absolute flux measurements in order
to minimize the observed systematic offset between the spectral and photometric magnitudes. 
This value matches the simulated slit-loss of a target slightly off-center from the 1$\arcsec$ slits with
a half light radius of $r_{h} = 0.18\arcsec$ in 0.8$\arcsec$ seeing (Figure 4 of Lem09), reasonable values 
given our observing conditions and the ACS \emph{F}814\emph{W} half-light radii of the galaxies observed.

\begin{figure}
\epsscale{1.2}
\plotone{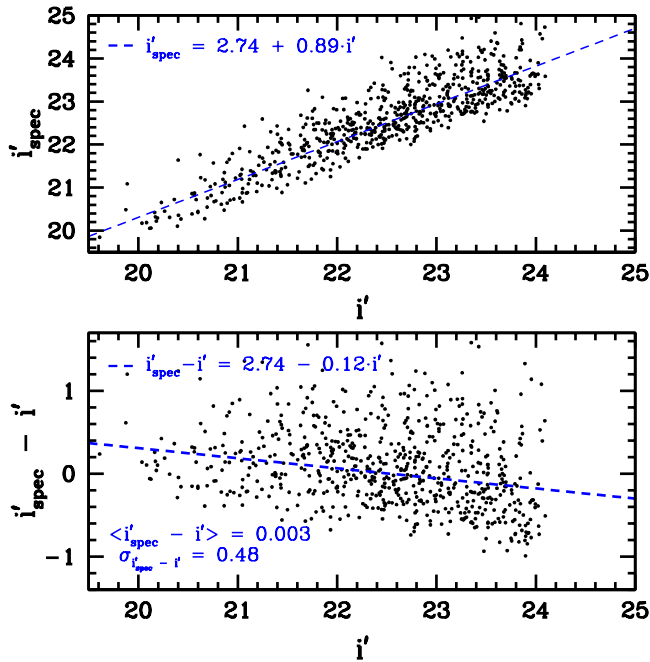}
\caption{Top Panel: LFC SDSS calibrated $i\arcmin$ magnitudes ($i\arcmin$) plotted as a function of
slit-loss corrected DEIMOS spectral $i\arcmin$ magnitudes ($i_{\rm{spec}}$, see \S4.2.1) for all members
in the Cl1604 supercluster with high quality (Q$\geq$3) spectra that fell near the middle of the slit and were not photometrically flagged. 
Bottom Panel: The difference in the spectral and LFC $i\arcmin$ magnitudes as a function of $i\arcmin$ for the same galaxies.
The best fit relations are overplotted. The large scatter in both panels represent real uncertainties 
in flux calibration of the data. An average slit throughput of $\omega_{\rm{slit}}=0.37$ was adopted for all 
spectra. This correction reproduces well the LFC $i\arcmin$ magnitudes on average. This throughput is optimized for the
average $i\arcmin$ magnitude of our NIRSPEC galaxies (i$\arcmin \sim$ 22.2) to avoid the magnitude dependent 
bias in the flux calibration that arises from the size-magnitude relationship.}
\label{fig:SDSS}
\end{figure}

The spectral magnitude, $i\arcmin_{\rm{spec}}$, is compared to our LFC photometry (see G08 
for details) in Figure \ref{fig:SDSS}. The rms scatter of the spectral magnitudes 
is 0.48 magnitudes, resulting in a $\sim$ 45\% uncertainty in any absolute flux
measurement. While there is little systematic bias, on average, between the spectral magnitudes 
and the LFC $i\arcmin$ magnitudes, there does exist a noticeable trend of decreasing 
$i\arcmin_{\rm{spec}}$-$i\arcmin$ with increasing $i\arcmin$ magnitude (see bottom panel of Figure \ref{fig:SDSS}). 
However, the NIRSPEC targets are primarily intermediate brightness cluster members 
($i\arcmin \sim 21.5-22.5$) and comparisons between the spectral and photometric magnitudes
at these magnitudes result in a distribution consistent with no systematic bias. We, 
therefore, ignore this bias and adopt an absolute uncertainty of 45\% in any DEIMOS flux measurements, 
resulting purely from the r.m.s.\ scatter in the measured magnitudes.

\subsubsection{NIRSPEC Flux Calibration}

The two standards observed with NIRSPEC, HD105601 (A2) and HD203856 (A0), were chosen from the
because of their similar spectral and luminosity class to $\alpha$Lyr and their low airmass at evening and
morning twilight. Analysis of the count rates of the two standard stars
resulted in observed variations of $\sim$20\% between the two nights, but stable
conditions during each individual night. Because HD203856 has an
identical spectral class to $\alpha$Lyr, absolute photometry was determined
by scaling the spectrum of $\alpha$Lyr (Colina et al.\ 1996) by the 
HD203856 Two Micron All Sky Survey (2MASS; Skrutskie et al.\ 2006) \emph{J}-band magnitude of 6.896 $\pm$ 0.023. The scaled $\alpha$Lyr spectrum was divided by the 
extinction-corrected composite spectrum of HD203856\footnote{Extinction correction was performed by correcting for the known \emph{J}-band 
atmospheric extinction on Mauna Kea extinction, http://www2.keck.hawaii.edu/inst/nirc/exts.html} created from
observations of the standard star taken on each night.

Our flux calibration method implicitly involves a slit-loss correction for a point
source (HD203856) under our observing conditions and will underpredict the  
slit-loss for a source with finite intrinsic angular extent. However, 
based on simulations similar to those performed in \S3.3.1, we find maximal 
slit-loss of only 10-15\% relative to a point source for galaxies with a 0.3$\arcsec$ 
half-light radius in our observing conditions. 
No correction is applied for this effect as we cannot accurately quantify the J-band 
half-light radius of each NIRSPEC target. Relative measurements (e.g., EWs or 
ratios of line fluxes) are unaffected by such losses. For absolute flux measurements any losses of this nature will result in 
underestimates of the true flux. The error associated with each line luminosity is then a quadrature sum of 
the random errors discussed in Appendix A and the systematic error associated with either 
the DEIMOS (for [OII] line luminosities) or NIRSPEC (for H$\alpha$ line luminosities) flux calibration. 

\subsubsection{Extinction Corrections} 

An additional uncertainty involved in comparing absolute flux measurements is due to internal extinction. 
Internal extinction corrections have been made using a variety of methodologies (e.g., Kennicutt 1983;
Kaufman et al.\ 1987; Wang \& Heckman 1996; Jansen et al.\ 2001; Kewley et al.\ 2002, 2004; Kauffmann et al.\ 2003a;
Buat et al.\ 2005; Moustakis et al.\ 2006; Rudnick et al.\ 2006; Weiner et al.\ 2007). While such methods have been shown to be
reliable for statistical samples of galaxies (though with differing levels of scatter, see Argence \& Lamareille 2009 
for a review), applying corrections to absolute line luminosities of an individual galaxy
may lead to significant biases. 

For this data, we attempt dust corrections based on (1) the luminosity of the H$\alpha$ line relative to the 24$\mu$m 
luminosity in those NIRSPEC targets detected in the MIPS data, (2) the absolute B-band magnitude, 
and (3) extinction values derived from spectral energy distribution (SED) fitting to broadband photometry 
($r\arcmin i\arcmin z\arcmin K_{s}+$IRAC). The details of each extinction correction method as well as the range of 
extinction values derived from each method are given in Appendix B. For all extinction corrections we assume a reddening curve, 
$k\arcmin(\lambda)$, parameterized by Calzetti et al.\ (2000), a total 
\emph{V}-band obscuration of $R\arcmin_{V} = 4.05 \pm 0.80$, and a value of $E_{S}(B-V) = (0.44 \pm 0.03)E(B-V)$, 
where $E_{S}(B-V)$ is the extinction of early-type stellar continua and $E(B-V)$ is the nebular 
extinction. The Calzetti reddening curve is adopted because it is valid over our range of extinction values and
is known to properly characterize the dust properties of galaxies at $z<1$ (Caputi et al.\ 2008; Conroy 2010). Further details regarding 
each extinction correction method as well as the range of extinction values derived from each method are given in Appendix B. 

Although the mean extinction values derived by the above three methods are consistent at the 1$\sigma$ level (see Appendix B), 
the $E(B-V)$ of an individual galaxy using each of the three methods frequently varied at levels exceeding 3$\sigma$. Therefore, when determining
intrinsic line luminosities from our NIRPSEC data we choose not to correct an individual galaxy using any of the above methods and instead
adopt a constant $E(B-V)$ = 0.3 for all galaxies. This value was chosen because it is consistent
with the observed mean extinctions found by all three methods (see Appendix B) and is well motivated from the observed extinctions 
in local samples
that span a large range of galaxy types and luminosities [Nearby Field Galaxies Survey (NFGS): Jansen 2000; Moustakas
\& Kennicutt 2006]. In \S5.4 we discuss the implications of choosing a constant extinction value rather than one of the
above three methods. Table \ref{tab:starformation} lists the observed fluxes and luminosities of the H$\alpha$ and [OII] lines of our
sample, as well as the extinction corrected luminosities.

\section{Results and Discussion}

The [OII] $\lambda$3727\AA\ doublet can originate from many processes that include, but are not limited to,
star formation, such as, e.g., galactic shocks (Heckman 1980; Dopita \& Sutherland 1995; Veilleux et al.\ 1995), cooling flows 
(Heckman 1981; Heckman et al.\ 1989), photoionization by hot stars (Terlevich \& Melnick 1985; Filippenko \& Terlevich 1992; Shields 1992), 
post-asymptotic giant branch stars (Binette et al.\ 1994; Taniguchi et al.\ 2000), or emission from an AGN 
(Ferland \& Netzer 1983; Filippenko \& Halpern 1984; Ho et al.\ 1993; Filippenko 2003; Kewley et al.\ 2006). While 
these processes can also result in residual levels of H$\alpha$ flux, in general, classification that relies on the strength of the 
H$\alpha$ feature has been shown to be much less sensitive to contamination by these processes (Kauffmann et al.\ 2003b; Y06). 
In this section we consider the relative strengths of [OII] and H$\alpha$, as well as the relative strengths
of [NII] and H$\alpha$ in order to probe the nature of the dominant source of ionizing flux in high-redshift cluster
galaxies. Following this, we investigate the other properties of galaxies dominated by LINER/Seyfert emission and 
those classified as star forming to determine their role in the context of galaxy evolution. 

\subsection{H$\alpha$ and [OII] Equivalent Width Properties of the NIRSPEC Sample}

\begin{figure*}
\plotone{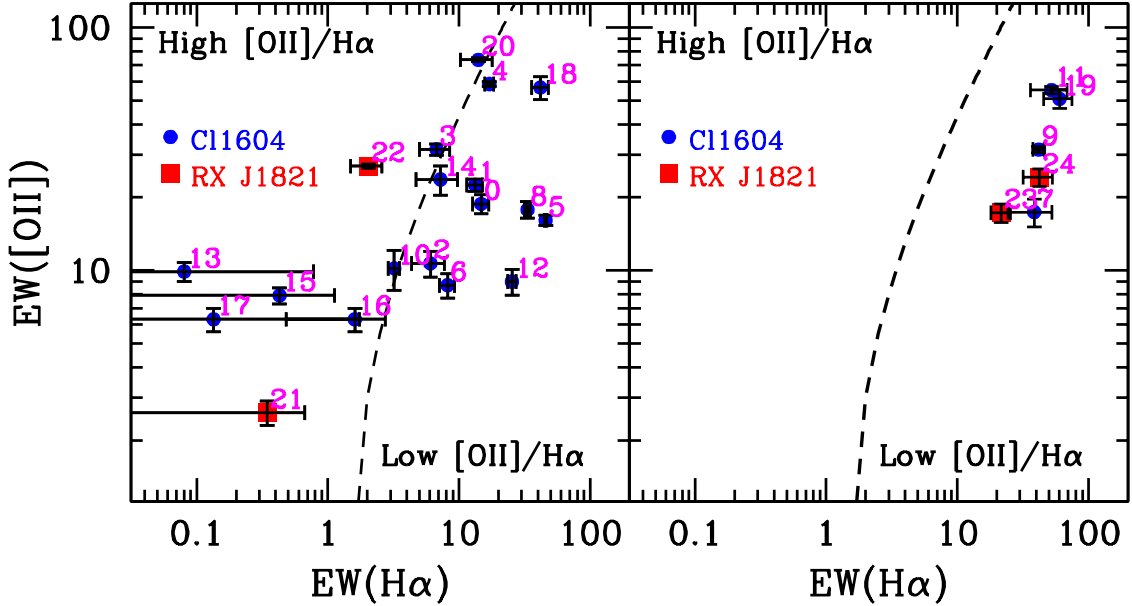}
\caption{ EW([OII]) vs. EW(H$\alpha$) for the 19 galaxies in our main sample (left) and the 
six galaxies in our filler sample (right). As discussed in \S4.1, we include the measured EWs of all galaxies 
regardless of the significance of the detection. All [OII] features have EWs that are detected in 
emission at $>3\sigma$. The dividing line between ``high-[OII]/H$\alpha$" and ``low-[OII]/H$\alpha$" adopted from the 
low-redshift sample of Y06 is plotted as a dashed line. While the main sample covers a large dynamic range in this space,
spanning four orders of magnitude in H$\alpha$ EW and approximately two orders of magnitude in [OII] EW, the filler
sample are all low-[OII]/H$\alpha$ galaxies and are confined to a narrow region in this phase space. Galaxy symbols are
coded by cluster membership and are labeled with galaxy numbers that correspond to the numbering in Tables 
\ref{tab:obs1}-\ref{tab:imaging}.}
\label{fig:EWEWraw}
\end{figure*}

\begin{figure}
\epsscale{1.2}
\plotone{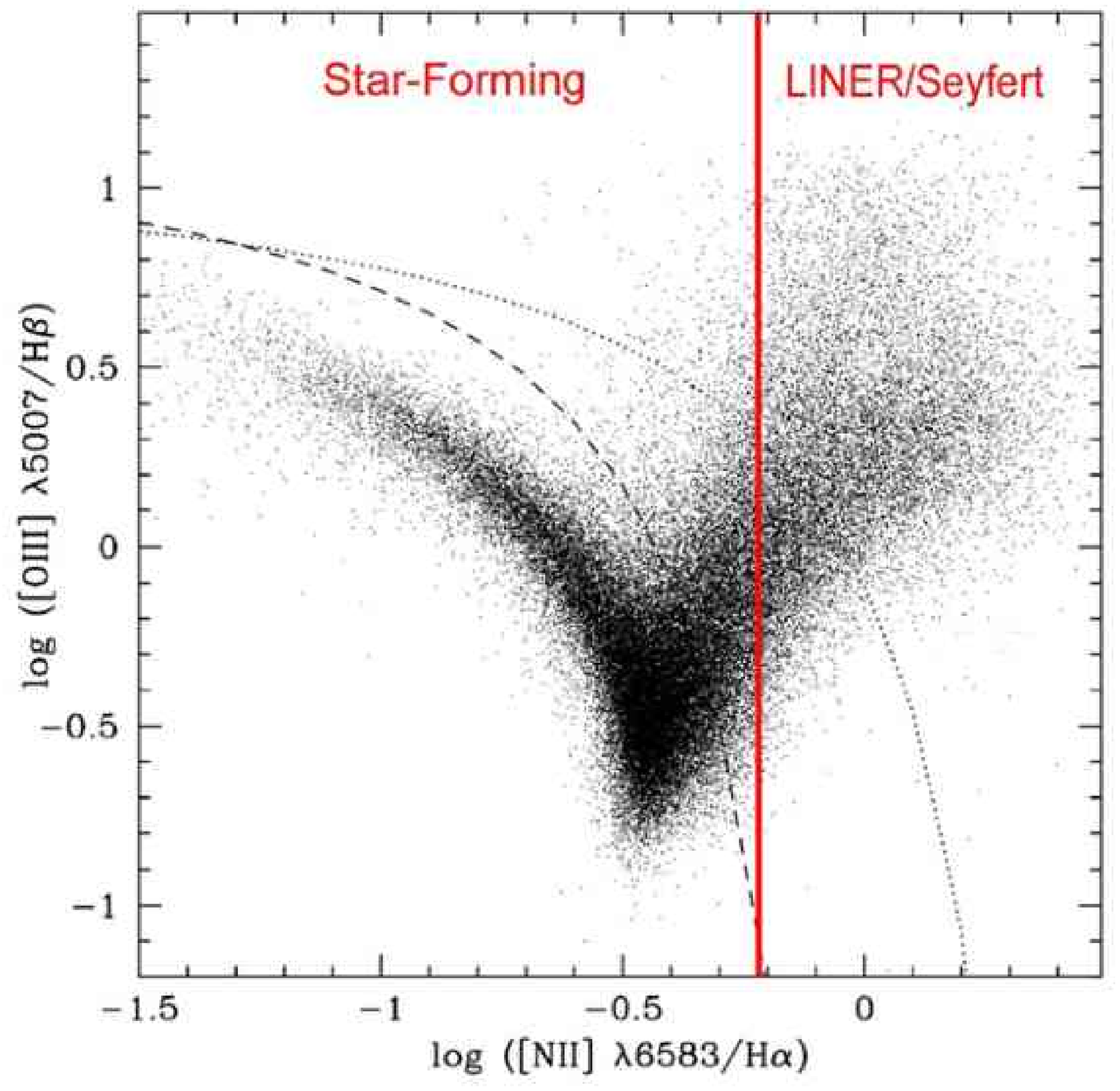}
\caption{Figure adopted from Kauffmann et al.\ (2003b) showing the emission line flux ratios for $\sim$55,000 SDSS galaxies used to discriminate the emission
class of each galaxy. The dashed line denotes the dividing line between star-forming galaxies and those with emission from a LINER or a Seyfert defined
by Kauffmann et al. The dotted line shows the maximum boundary of a pure starburst galaxy from theoretical modeling performed by Kewley et al.\ (2001)
(i.e., the ``extreme starburst line"). The region in between the two lines delineates galaxies with superpositions of
star formation and LINER/Seyfert activity (i.e., TOs). The
red solid vertical line shows our adopted boundary of log($F_{\rm{[NII]}/\rm{H}\alpha})=-0.22$ between star forming and LINER/Seyfert galaxies.
Note that this boundary sets only a lower limit to the number of LINERs, Seyferts, and TOs in our sample, as there exist a significant number of galaxies
in the SDSS sample that are inconsistent with pure star formation (mostly TOs) with -0.4 $<$ log($F_{\rm{[NII]}/\rm{H}\alpha}$) $<$ -0.22.}
\label{fig:Kauffmanndef}
\end{figure}

In Figure \ref{fig:EWEWraw} we plot the rest-frame EW of [OII] against the rest-frame
EW of H$\alpha$ for all 25 galaxies in our sample. Due to the small number of galaxies observed with NIRSPEC, we separate
populations that have [OII] EWs greater than expected, given the relative strength of H$\alpha$, from normal star-forming processes 
using the classification employed by Y06 for 55,000 galaxies at low redshift. The boundary, shown as a dashed line in 
Figure \ref{fig:EWEWraw}, is given by:

\begin{equation}
EW([OII]) = 5EW(H\alpha)-8;
\label{eqn:EWboundary} 
\end{equation}

\noindent Galaxies to the left of the line are classified as ``high-[OII]/H$\alpha$"; 
galaxies to the right of the line are classified as ``low-[OII]/H$\alpha$". In the low-redshift sample (Y06) 
there exist two populations that are well represented by log normal distributions, one centered at EW([OII]) $\approx$ 10\AA\ and 
EW(H$\alpha$) $\approx$ 14\AA\ (their low-[OII]/H$\alpha$ population) and one centered at EW([OII]) $\approx$ 6\AA\ and EW(H$\alpha$) 
$\approx$ 1\AA\ (their high-[OII]/H$\alpha$ population; see Figure 2 of Y06). The differences between these two populations at 
low redshift were generally due to the dominant emission mechanism present in each population. 
High-[OII]/H$\alpha$ and red low-[OII]/H$\alpha$ populations of Y06 generally had emission line ratios that were consistent with 
emission from a source other than star formation (i.e., a Seyfert or, more commonly, a LINER). Conversely, the blue 
low-[OII]/H$\alpha$ galaxies in this same study typically exhibited emission line ratios consistent with normal star formation. 
Table \ref{tab:spec} lists the EW class of each galaxy in our sample. 

\begin{figure*}
\epsscale{1.15}
\plottwo{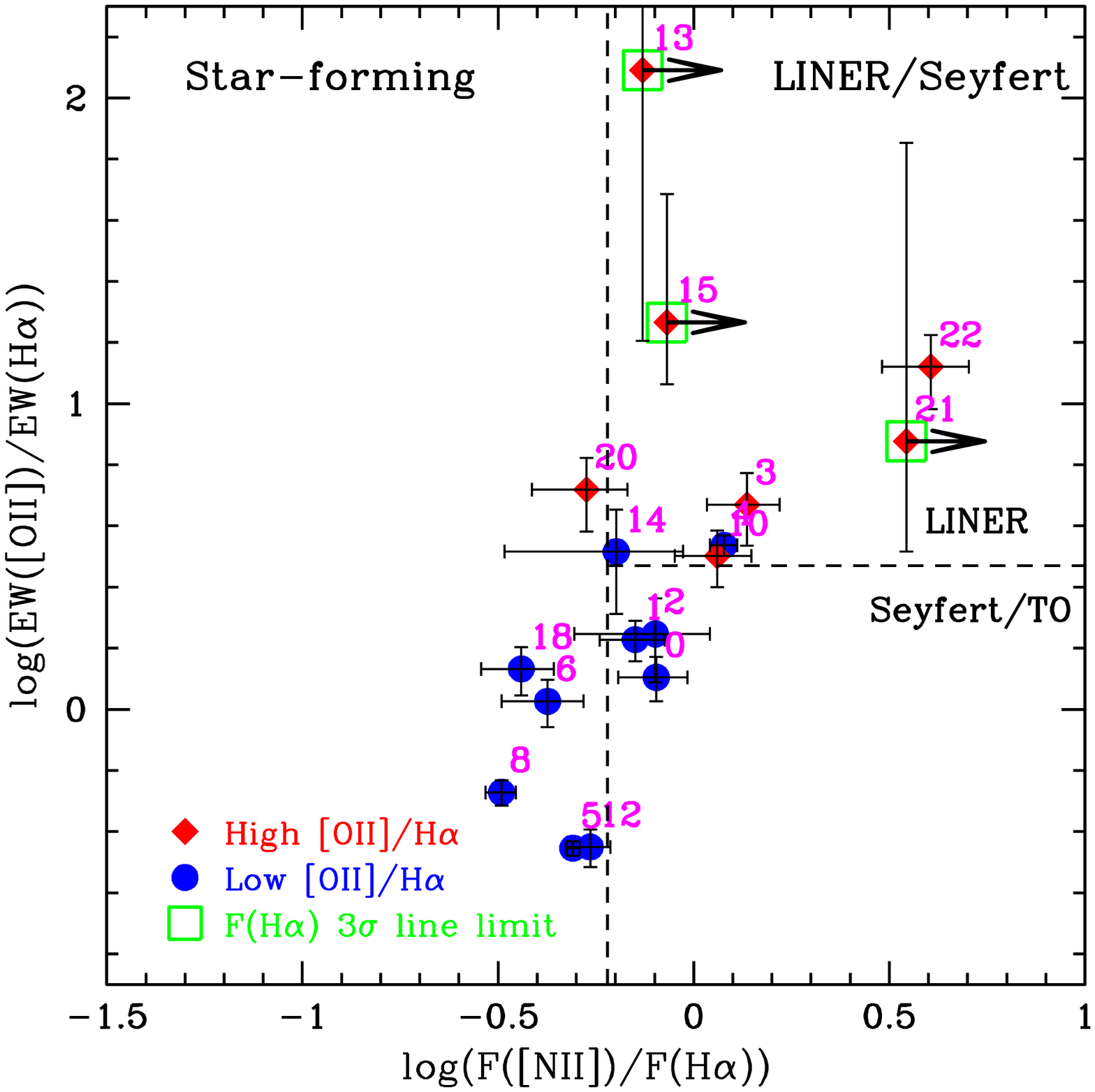}{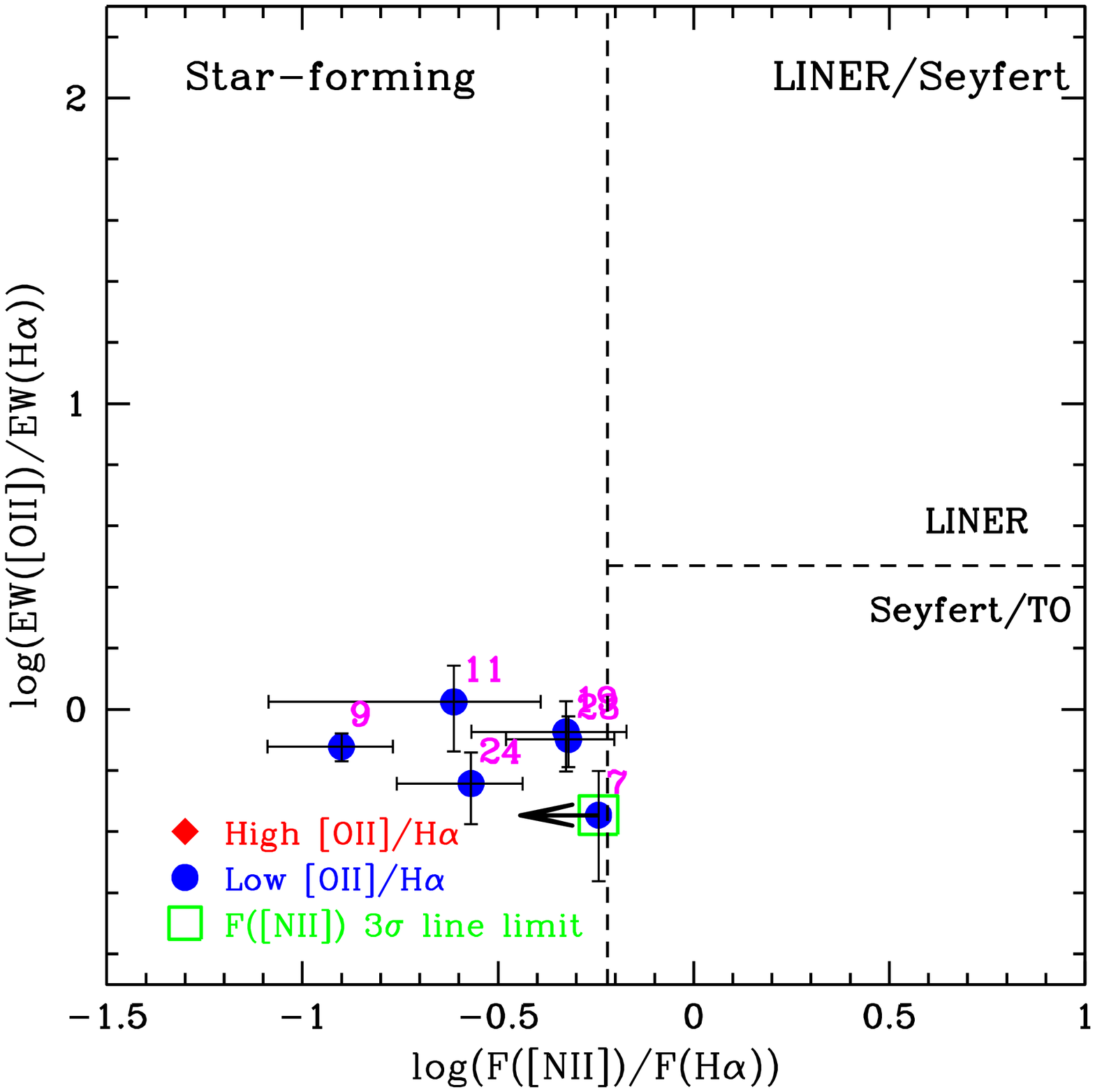}
\caption{(Left): Ratio of [OII] and H$\alpha$ EWs as a function of \emph{F}([NII])/\emph{F}(H$\alpha$) for 17 galaxies in the main
sample that have at least H$\alpha$ or [NII] detected at high ($>3\sigma$) significance. NIRSPEC targets in Cl1604 and RX J1821
are plotted. Galaxies with 3$\sigma$ upper limits on \emph{F}(H$\alpha$) are plotted with arrows. All galaxies have \emph{F}([NII]) detected at
$>3\sigma$. The vertical dashed line at log($F_{\rm{[NII]}/\rm{H}\alpha}=-0.22$) indicates our dividing line between star forming
and ``LINER/Seyfert" galaxies, adopted from the criterion used in Kauffmann et al.\ (2003b). The horizontal dashed line at 
log(EW([OII]/EW(H$\alpha$))=0.47 denotes our rough dividing line between LINERs and Seyfert/TOs (see \S5.2).  Galaxy labels are
identical to those in Figure \ref{fig:EWEWraw}. Eleven of the 17 galaxies plotted in the main sample are classified as LINER/Seyfert. (Right):
Same plot for the six galaxies in the filler sample, all of which are classified as star forming. The galaxy with a 3$\sigma$ limit
on \emph{F}([NII]) is plotted with an arrow. All galaxies have \emph{F}(H$\alpha$) detected at $>3\sigma$. There is a general trend of
increasing EW$_{\rm{[OII]}/\rm{H}\alpha}$ with increasing log($F_{\rm{[NII]}/\rm{H}\alpha}$).}
\label{fig:EWvslinerat}
\end{figure*}

Although this bimodality in our data is blurred by several galaxies at or near the boundary between low- and 
high-[OII]/H$\alpha$, the nine galaxies in our main sample that have moderate levels of [OII] emission
[EW([OII])$\la$ 10\AA] have H$\alpha$ EWs that span a factor of $\sim$300. This suggests that there exist significantly 
different mechanisms generating the ionizing flux in the galaxies in the main sample. The filler sample all belong
to the low-[OII]/H$\alpha$ population and cover a much smaller dynamic range in this space. This result is not surprising 
as these galaxies have DEIMOS spectra that are indicative of normal star formation processes. While our data suggest an 
interpretation similar to that of Y06, the mean absolute magnitude and 
rest-frame colors spanned by our sample differ appreciably from those in the Y06 sample. This difference may be important since 
the strength of the EW increases due to either (1) an increase in the luminosity of a spectral feature at a set continuum level, 
or (2) a fainter broadband magnitude at a fixed line luminosity. 

Differential changes in the stellar continuum levels can have a substantial effect on the observed correlations as 
the [OII] EW may be artificially inflated (in the sense of not being tied directly to the 
strength of a star formation episode) with respect to the H$\alpha$ EW in galaxies with
$g\arcmin-r\arcmin$ colors. However, if we adopt observed frame $i\arcmin-z\arcmin$ or $i\arcmin-K_{s}$ color as a 
proxy for the difference in continuum levels surrounding the [OII] and H$\alpha$ features, the Cl1604 
high-[OII]/H$\alpha$ population is only $\sim$0.1 mag redder than the low-[OII]/H$\alpha$ sample. Therefore, variations in 
stellar continua are not responsible for the observed difference between our
low-[OII]/H$\alpha$ and high-[OII]/H$\alpha$ populations; instead they must reflect
real variations in line luminosities. In the following sections we explore the nature of that variation.  

\subsection{Investigating the Dominant Source of Emission in the NIRSPEC Sample}

The three possible causes of variation in the observed luminosity of [OII] at a fixed 
H$\alpha$ value are 1) the primary source of ionizing flux, 2) extinction, and 3) changes in metallicity. 
To investigate the first possibility we utilize a modified version of the BPT diagrams (Baldwin, Phillips, \&
Terlevich 1981; Veilleux \& Osterbrock 1987). Traditionally, diagrams that utilize the strength of the H$\beta$ feature 
relative to [OIII] and the strength of the H$\alpha$ feature relative to a forbidden line (typically 6300\AA\ [OI], [NII], or 
6716\AA\ + 6731\AA\ [SII] due to their proximity to H$\alpha$) are favored (see Figure \ref{fig:Kauffmanndef}).
As we do not observe all the spectral features necessary for a full BPT analysis, we instead rely on 
a variation of this diagnostic to separate star-forming galaxies from LINERs and Seyferts.  We use [NII] 
rather than [OI] or [SII] as [OI] is typically too weak to detect, and [SII] coincides with bright OH airglow lines at 
$\lambda>1.25$\AA. In Figure \ref{fig:EWvslinerat} we plot our modified version of the 
BPT diagram for the main and filler samples. Only galaxies where [OII] and at least one other NIR emission line 
(H$\alpha$ or [NII]) were detected at high significance are plotted. 

Since we are collapsing the traditional \emph{F}([OIII])/\emph{F}(H$\beta$) (hereafter $F_{\rm{[OIII]}/\rm{H}\beta}$) ordinate of the BPT diagram, choosing a boundary 
along the \emph{F}([NII])/\emph{F}(H$\alpha$) (hereafter $F_{\rm{[NII]}/\rm{H}\alpha}$) axis that discriminates between galaxies dominated by emission 
from \ion{H}{2} regions and those dominated by a LINER or Seyfert is somewhat ambiguous. The large range of metallicities and ionization parameters present in star-forming 
galaxies results in a wide range of observed $F_{\rm{[NII]}/\rm{H}\alpha}$ values. A study of nearly 100,000 SDSS galaxies by
Kewley et al.\ (2006) found that this ratio varied from log($F_{\rm{[NII]}/\rm{H}\alpha}$)$\approx -1.5$ in extremely metal-poor star-forming galaxies to 
log($F_{\rm{[NII]}/\rm{H}\alpha}$)$\approx-0.3$ in those with super-solar abundances. Without the additional information that 
$F_{\rm{[OIII]}/\rm{H}\beta}$ provides we divide star-forming galaxies from those dominated by a LINER or Seyfert component (referred 
hereafter as ``LINER/Seyfert" galaxies) at log($F_{\rm{[NII]}/\rm{H}\alpha}$)$=$-0.22. This value reflects the maximal boundary of star-forming galaxies in the sample 
of Kauffmann et al.\ (2003b) (see vertical line in Figure \ref{fig:Kauffmanndef}) and is similar to cuts used in other studies when H$\beta$ and [OIII] are weak or unobservable 
(e.g., Miller et al.\ 2003; Stasi{\'n}ska et al.\ 2006). With the adoption of log($F_{\rm{[NII]}/\rm{H}\alpha}$)$>$ $-0.22$ as our bound for LINER/Seyferts, 
we also include in this cut ``Transition Objects" (hereafter TOs). These galaxies have ionizing flux that is a composite of emission from power-law sources, emission from old populations of
metal-rich stars, and emission from \ion{H}{2} regions (Ho et al.\ 1993; Kauffmann et al.\ 2003b; Kewley et al.\ 2006). Even though our 
cut will classify a galaxy as a LINER/Seyfert if as little as 20\% of its overall ionizing flux originates from a power-law source,
it is sufficient for our study that any such galaxy must have \emph{some} contribution from processes that are not star forming.

The choice of log($F_{\rm{[NII]}/\rm{H}\alpha}$)$<$ $-0.22$ as a bound for star-forming galaxies will result in some galaxies that are not truly star forming being classified as such.
This is especially true for galaxies with log($F_{\rm{[NII]}/\rm{H}\alpha}$)$>-0.35$, where the ``contamination" by power-law sources (primarily Seyferts and TOs) becomes 
significant (Stasi{\'n}ska et al.\ 2006; see also Figure \ref{fig:Kauffmanndef}). For our main sample (plotted in the left panel of Figure \ref{fig:EWvslinerat}), 
no galaxies have log($F_{\rm{[NII]}/\rm{H}\alpha}$) $<$ -0.5, suggesting that the contamination by LINER/Seyferts in galaxies classified as star forming is quite high. 
This conservative selection is necessary since we are attempting 
to show that [OII] emission in many of main sample galaxies cannot be solely attributed to star-forming processes. 
Metal-enriched starbursts with unusually low ionization parameters may exhibit 
log($F_{\rm{[NII]}/\rm{H}\alpha}$) as high as 0.2-0.3 (i.e., the ``extreme starburst line" in Figure \ref{fig:Kauffmanndef}; Kewley et al.\ 2001), 
thus contaminating our LINER/Seyfert sample; however, such galaxies are somewhat rare at $z\sim0.1$ and likely extremely so at $z\sim1$ 
(Tremonti et al.\ 2004; Kewley \& Ellison 2008; Liu et al.\ 2008; Lara-L{\'o}pez et al.\ 2009; P{\'e}rez-Montero et al.\ 2009). 

\begin{figure*}
\plotone{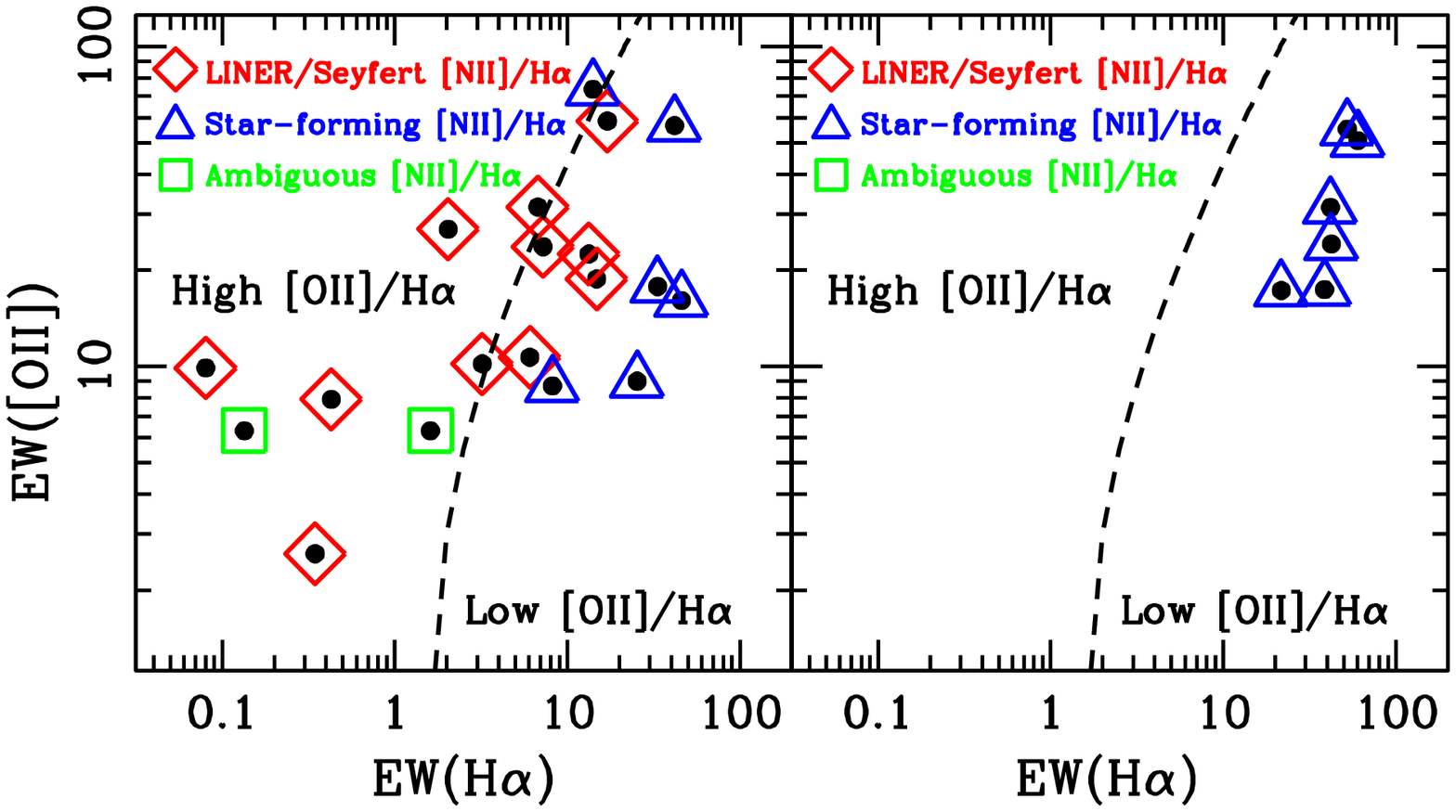}
\caption{Same as in Figure \ref{fig:EWEWraw}, EW([OII]) vs. EW(H$\alpha$) for the 19
galaxies in the main sample \emph{(left)} and the six galaxies in the filler sample \emph{(right)}. In this plot the error bars and galaxy
numbers are removed and replaced with the emission class of the galaxy. ``Ambiguous" is used for galaxies that do not have
detections in either H$\alpha$ or [NII] and thus cannot be definitively classified as either star forming or LINER/Seyfert;
however, it is likely that they belong to the latter class (see \S5.3). The division between high-[OII]/H$\alpha$ and
low-[OII]/H$\alpha$ is identical to that of Figure \ref{fig:EWEWraw}.
A large fraction ($\sim$62\%) of LINER/Seyfert and ambiguous galaxies are classified as high-[OII]/H$\alpha$. Galaxies
classified as star forming are almost exclusively low-[OII]/H$\alpha$. The one exception (galaxy 20) is likely a type 1.9
LINER/Seyfert (see \S5.2). All filler galaxies are classified as low-[OII]/H$\alpha$ and star forming.}
\label{fig:EWvsEWtype}
\end{figure*}

Of the 19 galaxies in our main (priority=1) sample, 11 ($\sim58$\%) are classified LINER/Seyfert, with several of the ``star-forming" galaxies lying within
1$\sigma$ of the dividing line. This result is significant, as priority 1 galaxies make up a large sub-section of the population in both Cl1604 and RX J1821. In
the absence of NIR spectroscopy all of the priority 1 galaxies would be considered star forming, significantly skewing the star-forming population 
to redder colors and brighter magnitudes (see Figure \ref{fig:priorityCMD}). This result will be discussed further in \S5.4. 

Two galaxies (16 and 17),
which we classify here as ``ambiguous" galaxies, did not have significant detections of either H$\alpha$ or [NII] despite having moderate [OII] emission [EW([OII])$\sim$7\AA]
and, thus, are not plotted in Figure \ref{fig:EWvslinerat}. These two
galaxies share many of the same emission, color, and morphological properties as our LINER/Seyferts.
The nature of the emission in these galaxies, which is likely due to a LINER or Seyfert, will be discussed further in \S5.3. 

Of the six galaxies comprising our filler (priority$\geq$2) sample, all have log($F_{\rm{[NII]}/\rm{H}\alpha}) \le -0.22$, 
consistent with pure star formation. These results include galaxies with significant detections in either H$\alpha$ or [NII] but no significant detection in
the other line (plotted in Figure \ref{fig:EWvslinerat} with arrows). In all such cases, the 3$\sigma$ upper
limit on the accompanying line was low enough to make a classification.

Though we cannot definitively distinguish between LINER, Seyfert, and TOs using $F_{\rm{[NII]}/\rm{H}\alpha}$ 
alone, the inclusion of EW([OII])/EW(H$\alpha$) (hereafter EW$_{\rm{[OII]}/\rm{H}\alpha}$) in our pseudo-BPT diagram allows for some distinction (i.e., Y06). LINERs exhibit typical EW$_{\rm{[OII]}/\rm{H}\alpha}$ values of $\sim$5 while Seyferts and TOs 
have significantly smaller ratios of $\sim$1. Though this ratio is somewhat sensitive to the metallicity of the host galaxy (see discussion in \S5.4), the 
average EW$_{\rm{[OII]}/\rm{H}\alpha}$ value of LINER/Seyfert and ambiguous galaxies observed in this sample (median EW$_{\rm{[OII]}/\rm{H}\alpha}$ = 3.9) strongly suggests 
this population is dominated by LINERs. The horizontal dashed line at log(EW$_{\rm{[OII]}/\rm{H}\alpha}$) = 0.47 in the two panels of Figure 
\ref{fig:EWvslinerat} provides a rough dividing line between LINERs and Seyferts/TOs.

As a conquence of this result, there is a noticeable trend between the ratio of [OII] and H$\alpha$ EWs and log(F$_{\rm{[NII]}/\rm{H}\alpha}$); 
galaxies generally exhibit higher EW$_{\rm{[OII]}/\rm{H}\alpha}$ ratios at higher values of 
log(F$_{\rm{[NII]}/\rm{H}\alpha}$). Of the 11 galaxies classified as LINER/Seyfert, six are high-[OII]/H$\alpha$. The trend of increasing LINER/Seyfert 
fraction with higher EW$_{\rm{[OII]}/\rm{H}\alpha}$ can be seen 
in Figure \ref{fig:EWvsEWtype}. This plot is identical to Figure \ref{fig:EWEWraw} except that 
the emission class of each galaxy is now indicated. All galaxies classified as star forming with one exception (galaxy 20) are significantly separated from 
the low-high EW$_{\rm{[OII]}/\rm{H}\alpha}$ dividing line. LINER/Seyferts cover a larger range in this phase space, but almost all lie nearly on 
or to the left of dividing line. Tables \ref{tab:mainvsfillertable} and \ref{tab:maintable} list the total number of galaxies in both 
the main and filler sample and how these samples break down as a function of EW ratios and emission classes. 

The notable exception to the trend of increasing LINER/Seyfert fraction with higher EW$_{\rm{[OII]}/\rm{H}\alpha}$ is galaxy 20. This galaxy is classified by its 
log($F_{\rm{[NII]}/\rm{H}\alpha}$) ratio as star forming but exhibits high levels of [OII] relative to H$\alpha$. Both the [OII] emission and H$\alpha$ in this 
galaxy appear slightly broadened ($\sim$300 km s$^{-1}$), typical of a type 
1.9 LINER/Seyfert. The broad line features likely skew the log(F$_{\rm{[NII]}/\rm{H}\alpha}$) ratio to a lower value, making the flux ratio difficult to interpret. 
Accordingly, we exclude this galaxy from any further analysis that compares the global properties of the LINER/Seyfert and star-forming samples.

Since the two ambiguous galaxies likely belong to the LINER/Seyfert class (albeit at a lower luminosity), the purity of the sample to the left of the dividing line 
suggests that at least a sub-population of LINER/Seyfert galaxies can be selected in high-redshift samples using only relative measurements of the [OII] and 
H$\alpha$ lines. This selection is important because it does not rely on the detection of the (usually) fainter [NII], [SII], or [OI] lines or on absolute spectrophotometry. 
Galaxies that are dominated by LINER emission (and not Seyfert or TO-like emission) can likely be selected in this manner without much loss in completeness or purity 
(see Figures \ref{fig:EWvslinerat} and \ref{fig:EWvsEWtype}). Low-[OII]/H$\alpha$ galaxies with LINER/Seyfert emission ratios, primarily Seyferts and TOs,  
will not be separable from star-forming galaxies using such a cut and must be selected, in the absence of optical forbidden lines, using other methods 
(e.g., X-ray emission, IR colors, radio power). 

\begin{deluxetable}{lcc}
\tablecaption{Overview and Classifications of NIRSPEC Targets \label{tab:mainvsfillertable}} 
\tablehead{ \colhead{} & \colhead{Main Sample} & \colhead{Filler Sample}}
\startdata
Total number & 19 & 6 \\
Cl1604 members& 17 & 4 \\
RXJ1821 members& 2 & 2 \\
High-[OII]/H$\alpha$ & 9 (47.4\%) & 0 \\
Low-[OII]/H$\alpha$ & 10 (52.6\%) & 6 (100\%)\\
LINER/Seyfert & 11 (57.9\%) & 0 \\
Ambiguous & 2 (10.5\%) & 0 \\
Star forming & 6 (31.6\%) & 6 (100\%)
\enddata
\end{deluxetable}

\begin{deluxetable}{lcc}
\tablecaption{Classifications of the NIRSPEC Main Sample \label{tab:maintable}}
\tablehead{ \colhead{} & \colhead{High-[OII]/H$\alpha$} & \colhead{Low-[OII]/H$\alpha$}}
\startdata
LINER/Seyfert & 6 (67\%) & 5 (50\%) \\
Ambiguous & 2 (22\%) & 0  \\
Star forming & ~~1 (11\%)\tablenote{The one high-[OII]/H$\alpha$ galaxy classified as star forming is likely a type 1.9 LINER/Seyfert, see \S5.2.} & ~5 (50\%)

\enddata
\end{deluxetable}

\subsection{Color, Spatial, and Morphological Properties of the NIRSPEC Sample}

\begin{figure*}
\epsscale{1.10}
\plottwo{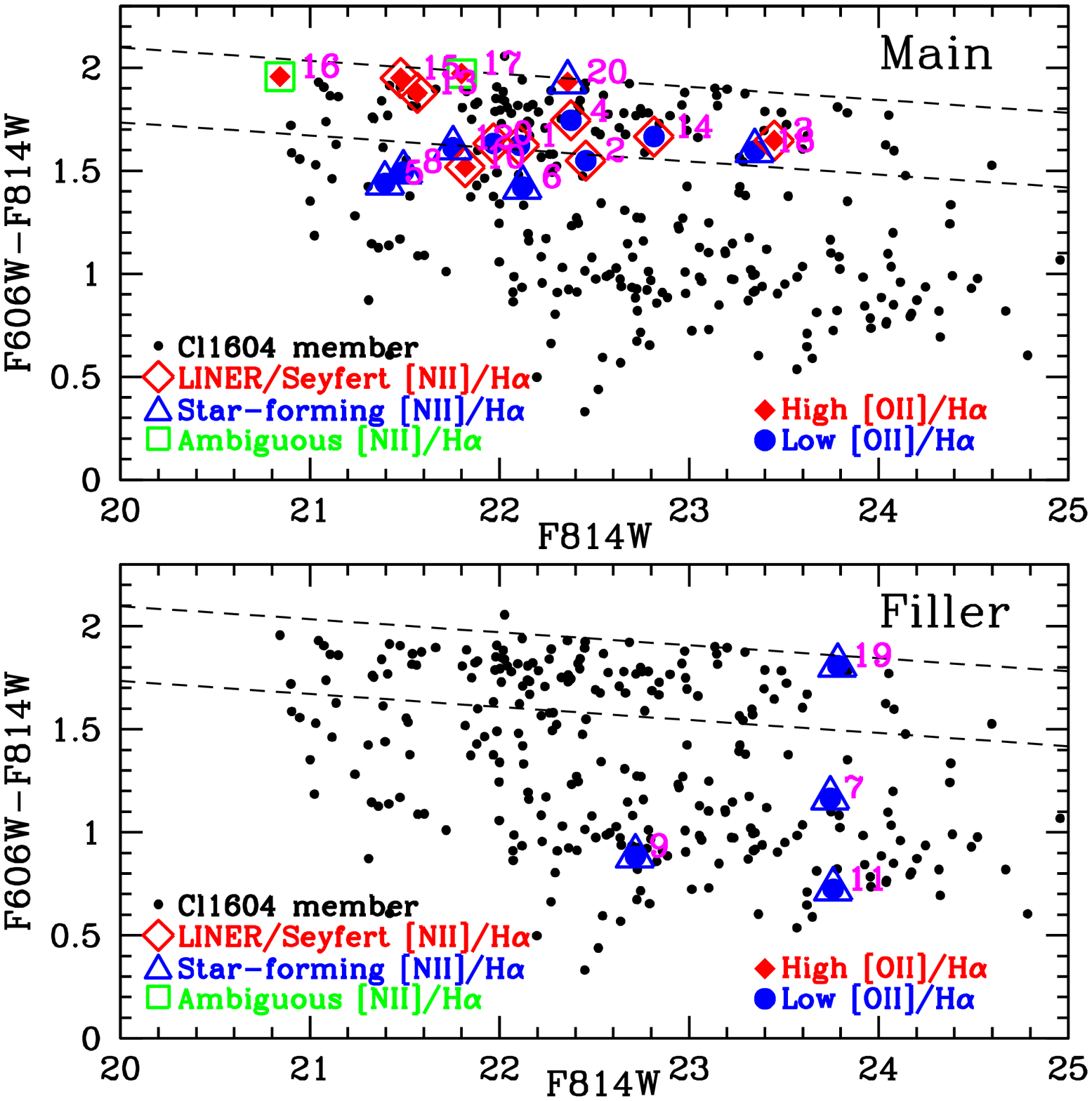}{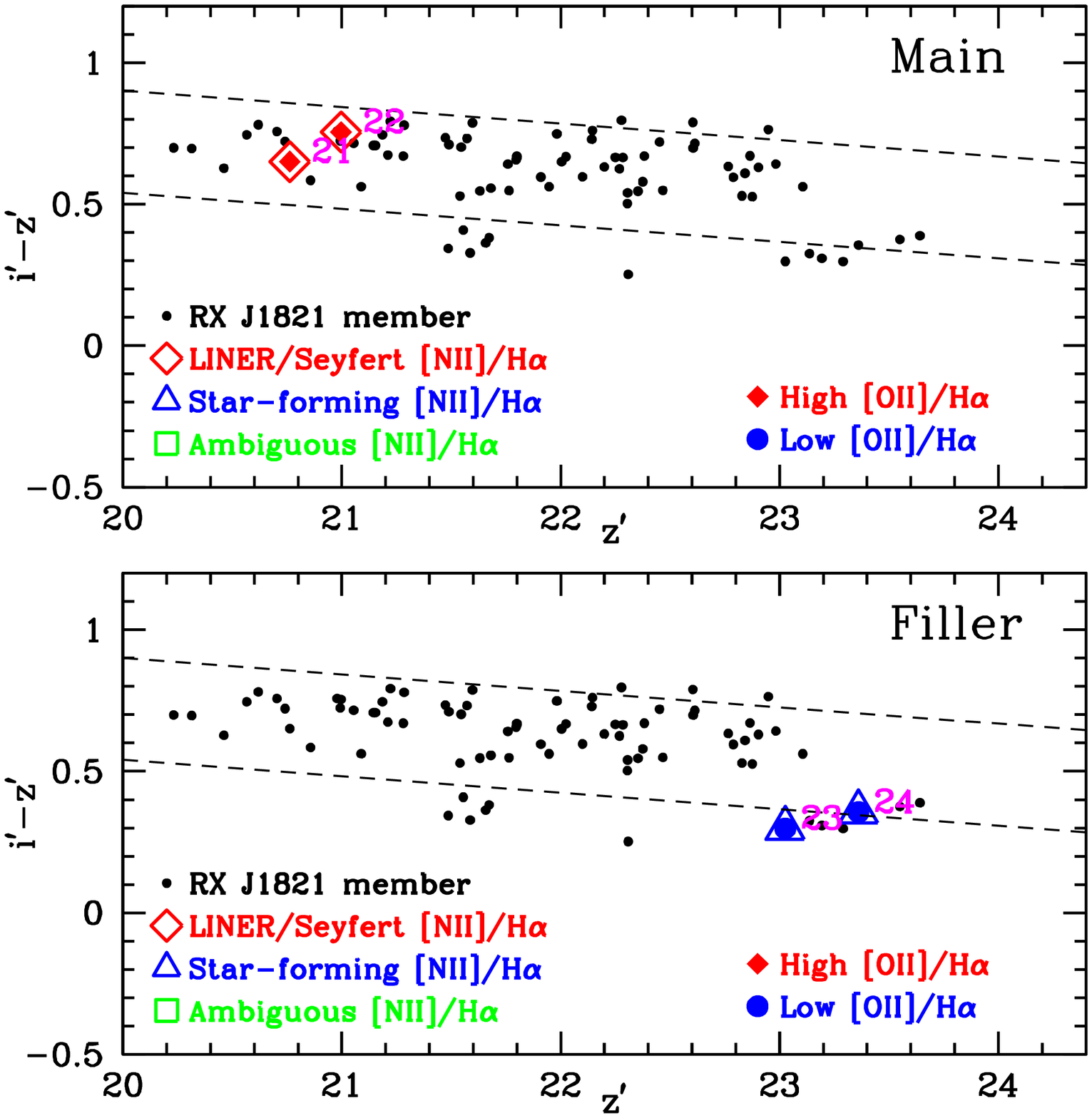}
\caption{(Left): ACS \emph{F}606\emph{W}-\emph{F}814\emph{W} color-magnitude diagram (CMD) for the 371 Cl1604 supercluster members observed in the ACS imaging. Cluster members not
observed with NIRSPEC are represented by small points. The 19 galaxies in the main sample (top panels) and the six galaxies in the filler sample
(bottom panels) observed with NIRSPEC are indicated by larger symbols. The dashed lines in the left panels denote the Cl1604 red sequence (see \S5.3).
The galaxy labels are the same as those in Figures \ref{fig:EWEWraw} and \ref{fig:EWvslinerat}. (Right): LFC $i\arcmin$-$z\arcmin$
CMD for the 72 confirmed members in RX J1821. The dashed lines denote the RX J1821 red sequence (see \S5.3). Note the significantly higher
fraction of members on the red sequence as compared to Cl1604. In total, 9 of the 13 red sequence galaxies
in the main sample have $F_{\rm{[NII]}/\rm{H}\alpha}$ indicative of a LINER/Seyfert and an additional two (galaxies 16 and 17) have other properties
that are highly suggestive of a LINER/Seyfert. A majority of red-sequence galaxies in the main sample also have high-[OII]/H$\alpha$. These results
suggest that many galaxies on the red sequence at high redshift are dominated by a LINER or Seyfert and are emitting [OII] as a consequence.
Many of the low-[OII]/H$\alpha$ LINER/Seyfert galaxies in Cl1604 lie at the blue end of the Cl1604 red sequence, possibly suggesting that these galaxies
are transitioning to the red sequence.}
\label{fig:NIRSPECCMD}
\end{figure*}

\begin{figure*}
\epsscale{0.60}
\plotone{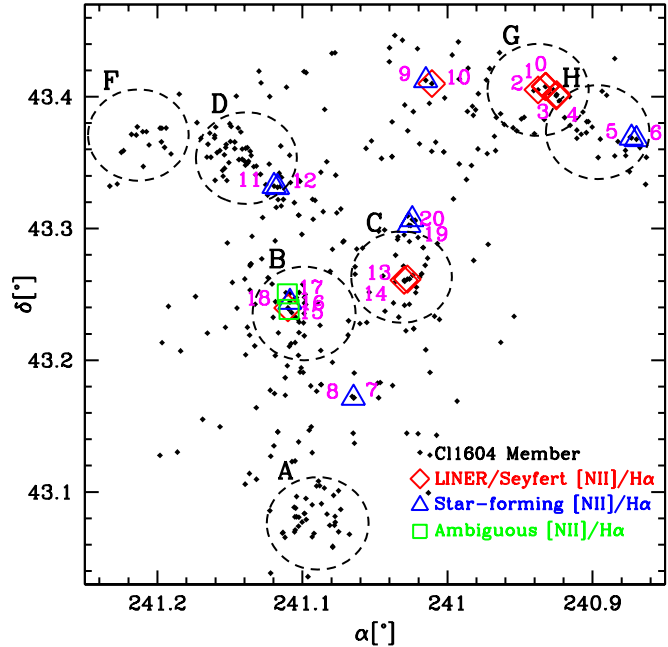}
\caption{Spatial distribution of our Cl1604 targets plotted with the other 393 confirmed Cl1604 supercluster
members. The letters for each constituent cluster or group are adopted from G08. The radius of each dashed line
represents the angular extent of 1 h$^{-1}$ Mpc at the redshift of each group/cluster. Generally, the LINER/Seyfert and ambiguous galaxies lie at
the centers of clusters and groups in the system. The structures that contain LINER/Seyfert galaxies extend over a large range
in mass, from 313 km s$^{-1}$ (cluster C) to 811 km s$^{-1}$ (cluster B) (G08). Star forming galaxies are generally found in
the connecting filaments.}
\label{fig:RAdectype}
\end{figure*}

We now investigate the color, spatial, and morphology distribution of our NIRSPEC sample. As shown in Figures \ref{fig:NIRSPECCMD} and 
\ref{fig:RAdectype}, the galaxies in our sample range
from faint blue-cloud galaxies in filaments between clusters (e.g., galaxies 7, 9, and 11) to the very reddest, most luminous galaxies 
at the center of well-established clusters (e.g., galaxies 13, 15, 16, and 17). Table \ref{tab:imaging} lists the 
$r\arcmin i\arcmin z\arcmin K_{s}$ magnitudes for all RX J1821 and Cl1604 galaxies observed with NIRSPEC, as well as the ACS 
magnitudes and morphologies for our Cl1604 sample. 

At low redshift the dominant emission mechanism in red galaxies with appreciable [OII] is considerably different than in blue galaxies
(Y06). Galaxies with high levels of [OII] relative to 
H$\alpha$ are almost exclusively red and classed as LINERs. Conversely, blue galaxies with low EW$_{\rm{[OII]}/\rm{H}\alpha}$ are predominantly 
star-forming galaxies. While a non-negligible fraction of red galaxies have low EW$_{\rm{[OII]}/\rm{H}\alpha}$, these galaxies show a 
different F$_{\rm{[NII]}/\rm{H}\alpha}$ distribution than their blue counterparts, typically being classed as Seyferts or TOs. 

In order to differentiate between galaxies of different colors in our own data, we
use simple color-magnitude cuts to separate red-sequence galaxies from bluer members in both structures. These color cuts 
were originally done in observed $i\arcmin$-$z\arcmin$, roughly equivalent at $z=0.88$ to  
SDSS $^{0.1}$($g\arcmin-r\arcmin$) cuts used by Y06 to differentiate between red and blue galaxies. In the Cl1604 field we define the red sequence in terms of the 
\emph{F}606\emph{W}-\emph{F}814\emph{W} colors rather than LFC $i\arcmin$-$z\arcmin$. The red sequence 
is defined using the observed colors of all confirmed members in the magnitude range 21 $\leq$ \emph{F}814\emph{W} $\leq$ 23 for Cl1604 and 20 $\leq$ $z\arcmin$ $\leq$ 22.5 
for RX J1821. For each field we fit the observed color distribution of the cluster galaxies with a linear function and subtract
the best-fit color-magnitude relationship from the observed color distribution defining a ``residual" color. For the two structures the best-fit
relationships are:

\begin{align}
\rm{F606W}-\rm{F814W} &= 3.182 - 0.063 \,\rm{F814W} ~~~\it{Cl1604}
\label{redsequence1604}\\
\nonumber\\
i\arcmin-z\arcmin &= 1.887 - 0.058 \, z\arcmin ~~~~~~~\it{RX J1821}
\label{redsequence5281}
\end{align}

\noindent Following standard methodology of defining the red-sequence in low to moderate redshift clusters (Gladders et al.\ 1998; Stott et al.\ 2009), we adopt 
the 3$\sigma$ scatter in the residual colors ($\sigma$=0.06) as the extent of the red sequence in color space for RX J1821.
In Cl1604 we adopt 2$\sigma$ for the width of the red-sequence, as the r.m.s. of the colors is much larger ($\sigma$=0.0907) due to the extent of 
the supercluster in redshift space. Of the 25 galaxies in our sample, 60\% (15 of 25) lie on the red sequence of the two structures, increasing to 
$\sim$70\% (13 of 19) in our main sample (see Table \ref{tab:colortable}). Ten galaxies (six in the main sample and four in the filler sample) have colors that are blueward 
of the red sequence.  

The high-[OII]/H$\alpha$ sample consists of 8 of the 13 red-sequence galaxies in the main sample and only one galaxy in the blue cloud (galaxy 10),
similar to the observed properties of high-[OII]/H$\alpha$ emitters at low redshift. The low-[OII]/H$\alpha$ population includes five of the six galaxies 
in the main sample that are blueward of the red sequence. This population has similar emission 
mechanisms to the blue-cloud low-[OII]/H$\alpha$ galaxies at low redshift, as 80\% of the blue low-[OII]/H$\alpha$ galaxies in the main sample have $F_{\rm{[NII]}/\rm{H}\alpha}$ 
values consistent with star formation. The one low-[OII]/H$\alpha$ blue-cloud galaxy in the main sample that is not consistent with pure star formation (galaxy 2) lies 
extremely close to the red sequence.

While blue low-[OII]/H$\alpha$ galaxies are primarily star forming, four of the five red-sequence low-[OII]/H$\alpha$ galaxies are classified as LINER/Seyfert.
However, while these galaxies are classified as LINER/Seyfert, their $F_{\rm{[NII]}/\rm{H}\alpha}$ values lie at the low end of the LINER/Seyfert distribution.
Of all galaxies classified as LINER/Seyfert where both [NII] and H$\alpha$ are detected with high significance,
those with low-[OII]/H$\alpha$ values on or near the red sequence comprise five of the six lowest $F_{\rm{[NII]}/\rm{H}\alpha}$ values.
As $F_{\rm{[NII]}/\rm{H}\alpha}$ decreases in LINER/Seyfert galaxies, the fractional contribution to the ionizing flux from the LINER/Seyfert component 
decreases monotonically (see Kewley et al.\ 2001 for a detailed discussion). Therefore, it is likely that low-[OII]/H$\alpha$ LINER/Seyfert galaxies have the least 
dominant LINER/Seyfert components and probably contain some residual star formation. Figure \ref{fig:EWvslinerat} confirms this result, as more than half of the 
low-[OII]/H$\alpha$ LINER/Seyferts lie below the line differentiating LINERs and Seyferts/TOs. The distributions of LINER/Seyfert and star-forming galaxies by color and 
EW ratio properties for the main sample are given in Table \ref{tab:emissioncolortable}.

Two of the eight red-sequence high-[OII]/H$\alpha$ galaxies were classified as ambiguous. These two galaxies (16 and 17)
are similar to galaxies classified as LINER/Seyfert, both in their high levels of [OII] emission relative to their formal limits on H$\alpha$ 
and in their color and spatial distributions. Indeed, almost all of the luminous red galaxies are classified as LINER/Seyfert and more than half 
have high levels of [OII] relative to H$\alpha$, suggesting that these types of galaxies are typically dominated by a LINER or a Seyfert. Since five of the six 
red-sequence galaxies with high-[OII]/H$\alpha$ and well-defined $F_{\rm{[NII]}/\rm{H}\alpha}$ are classified as LINER/Seyfert (the one exception is galaxy 20, which is 
itself likely an AGN, see \S5.2), ambiguous galaxies are also likely dominated by the same emission source. Based on these properties we include these 
ambiguous galaxies in our sample of LINER/Seyfert galaxies.

\begin{deluxetable}{lcc}
\tablecaption{Properties of the NIRSPEC Main Sample by Color \label{tab:colortable}}
\tablehead{ \colhead{} & \colhead{Red} & \colhead{Blue}  }
\startdata
Total & 13 & 6 \\
High-[OII]/H$\alpha$ & 8 (62\%) & 1 (17\%) \\
Low-[OII]/H$\alpha$ & 5 (38\%) & 5 (83\%) \\
LINER/Seyfert & 11 (85\%)\tablenote{The two ambiguous galaxies are included here in the LINER/Seyfert category, see \S5.3.} & 2 (33\%) \\
Star forming & 2 (15\%)\tablenote{One galaxy (20) classified as star forming is likely a type 1.9 LINER/Seyfert, see \S5.2.} & 4 (67\%) \\
Early type & 8 (73\%)\tablenote{Only 11 of the 13 red-sequence galaxies in the main sample have morphological information.} & 2 (33\%) \\
Late type & 3 (27\%)\tablenotemark{c} & 4 (67\%)

\enddata
\end{deluxetable}

\begin{deluxetable*}{lcccc}
\tablecaption{Properties of the NIRSPEC Main Sample by Emission Class and Color \label{tab:emissioncolortable}}
\tablehead{ \colhead{} & \colhead{~~~~~~~~~~~~~~~~~~~~~~~Red} & \colhead{} & \colhead{~~~~~~~~~~~~~~~~~~~~~~~Blue~} & \colhead{}}
\startdata
\sidehead{ ~~~~~~~~~~~~~~~~~~~~~~~~~~~ High-[OII]/H$\alpha$ ~~ Low-[OII]/H$\alpha$ ~ High-[OII]/H$\alpha$ ~~~~~ Low-[OII]/H$\alpha$ } \hline \\[-1ex]
LINER/Seyfert & 7 (54\%)\tablenote{The two ambiguous galaxies are included here in the LINER/Seyfert category, see \S5.3.} & 4 (30\%) &  1 (8\%) & 1 (8\%) \\
Star forming & 1 (17\%)\tablenote{One galaxy (20) classified as star forming is likely a type 1.9 LINER/Seyfert, \S5.2.} & 1 (17\%) & 0 & 4 (66\%)
\enddata
\end{deluxetable*}

The \emph{F}606\emph{W} and \emph{F}814\emph{W} images of each galaxy in the Cl1604 field were visually inspected by one of the authors (LML) in order to classify morphologies 
(galaxies in RX J1821 are excluded from this analysis due to the lack of ACS data). The morphologies of our sample show a spread in properties similar to 
that of their color and environments, ranging from irregular Sc galaxies (5 and 8) to isolated elliptical galaxies (3, 16, and 20). 
Table \ref{tab:imaging} lists the morphologies of all our targets. The nine galaxies classified as LINER/Seyfert in Cl1604, along with the two 
galaxies classified as ambiguous, are almost all early-type galaxies, further supporting our claim that the ambiguous galaxies are similar to the LINER/Seyfert population.  
The exceptions are galaxies 0 and 1, both morphologically classified as disk galaxies, although clearly disturbed by a merger or interaction. In total 
six galaxies are classified as star forming in the Cl1604 main sample. This population is dominated by late-type morphologies, with four of the six galaxies having 
morphologies consistent with either spirals or amorphous galaxies. The four galaxies in the Cl1604 filler sample, all classified as star forming, also had late-type morphologies.
Automated measurements of galaxy compactness are similarly disparate between the two emission classes, as the LINER/Seyfert and ambiguous galaxies are, on average,  
more compact than the star-forming galaxies at the 98\% CL. Table \ref{tab:colortable} lists the distribution of EW ratios, emission classes, and morphological types
as a function of color for the main sample. 

\subsection{Understanding the Nature of [OII] Emission in LINER/Seyfert Galaxies}

\begin{figure}
\epsscale{1.2}
\plotone{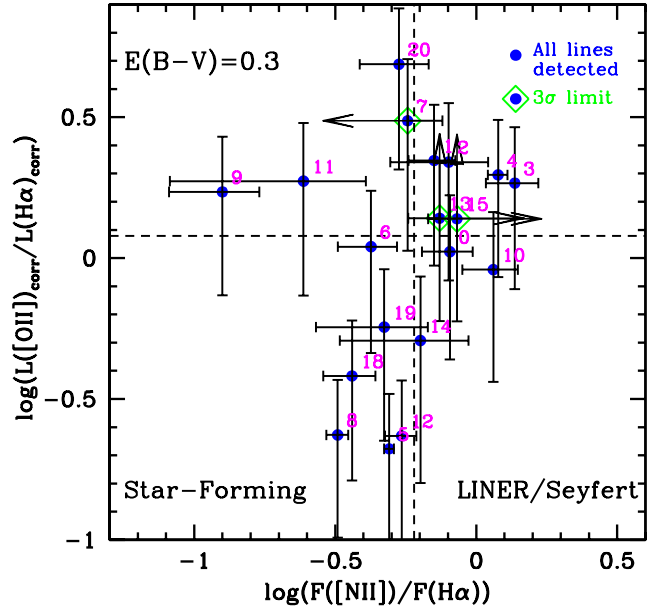}
\caption{Logarithm of the ratio of extinction corrected \emph{L}([OII]) and \emph{L}(H$\alpha$) as a function of log($F_{\rm{[NII]}/\rm{H}\alpha}$) for 
the Cl1604 galaxies in our sample. The extinction correction is performed
using a constant $E(B-V)=0.3$ and the Calzetti et al.\ (2000) reddening law for all galaxies. Galaxies 16 and 17, which
do not have significant detections in either H$\alpha$ or [NII], are omitted from this plot. Galaxies with 
significant detections in either H$\alpha$ or [NII], but with no detection in the accompanying line have 3$\sigma$ upper 
limits plotted as arrows. The vertical dashed 
line at $F_{[NII]/\rm{H}\alpha}$=-0.22. denotes our boundary between a star forming and a LINER/Seyfert classification adapted from Kauffmann et al.\ (2003b). 
The horizontal dashed line [log($L_{\rm{[OII]}/\rm{H}\alpha}$)=0.08] is the average extinction corrected luminosity ratio for star-forming galaxies at 
low-redshift (Kewley et al.\ 2004). Almost all of the galaxies classified as LINER/Seyfert lie above this line, while many of the  
star-forming galaxies lie below. This result suggests that the LINER/Seyfert component contributes appreciably to the [OII] 
emission. There is a general trend of increasing [OII] emission relative to H$\alpha$ with increasing $F_{\rm{[NII]}/\rm{H}\alpha}$, with a few 
notable exceptions.}
\label{fig:Lumratiovslineratio}
\end{figure}

While the high levels of [OII] relative to H$\alpha$ correlate well with the presence of a LINER/Seyfert in a galaxy, it may still be the case  
that the [OII] emission in such galaxies is due to metallicity or extinction effects. In 
Figure \ref{fig:Lumratiovslineratio} we plot the ratio of the extinction corrected [OII] and H$\alpha$ luminosities (hereafter $L_{\rm{[OII]}/\rm{H}\alpha}$)
vs. $F_{\rm{[NII]}/\rm{H}\alpha}$ for the 19 Cl1604 targets with significant detections in [OII] and at least one of the two other spectral features 
(H$\alpha$ or [NII]). While there are significant exceptions (e.g., galaxies 9, 10, 11, and 20), there is a general trend of increasing [OII] luminosity relative 
to H$\alpha$ for galaxies with higher levels of [NII] relative to H$\alpha$. Excluding galaxy 20 (likely a type 1.9 LINER/Seyfert),  
a Spearman rank correlation coefficient test on our Cl1604 NIRSPEC sample results in a probability of positive correlation between the two ratios 
of 83\%, increasing to $>$99.99\% when only the 14 galaxies in the main sample are used. Line luminosities are extinction corrected using a constant value of $E(B-V)$=0.3 
rather than individual extinction corrections based on the techniques discussed in \S4.2.3 and Appendix B, as the overall trend (and Spearman rank coefficient) remains virtually 
unchanged regardless of our choice of extinction correction. Thus, the observed excess [OII] emission in LINER/Seyfert galaxies is not likely due to dust effects. 

Excluding galaxy 20 from the star-forming sample (since it is likely a type 1.9 LINER/Seyfert), the mean $L_{\rm{[OII]}/\rm{H}\alpha}$ ratio
for star-forming galaxies is 1.04$\pm$0.27, consistent with the average extinction corrected $L_{\rm{[OII]}/\rm{H}\alpha}$ of 1.2 (shown 
as a dashed line in Figure \ref{fig:Lumratiovslineratio}) found for the NFGS sample analyzed by Kewley et al.\ (2004). 
The \emph{lower limit} on the average corrected $L_{\rm{[OII]}/\rm{H}\alpha}$ ratio for all galaxies classified as LINER/Seyferts is 1.51 $\pm$0.28, higher than 
both that of our star-forming galaxies and the low-redshift NFGS sample. This mean is a lower limit because several of the LINER/Seyfert galaxies 
in our sample have 3$\sigma$ upper limits on their H$\alpha$ luminosities. Removing these galaxies from our sample increases the mean 
$L_{\rm{[OII]}/\rm{H}\alpha}$ for LINER/Seyfert galaxies, confirming the significance of this result.

This ratio, however, can be strongly affected by galaxy metallicity. The effect of increasing metallicity on this ratio for galaxies with $Z\ga 0.6Z_{\odot}$ 
is to \emph{decrease} the intrinsic $L_{\rm{[OII]}/\rm{H}\alpha}$. The average metallicity of the low-redshift NFGS sample is just above solar 
[$\langle$log(O/H)+12$\rangle$ $\sim$ 8.75], using the metallicity calibration of Kewley \& Dopita (2002). If our targets are, on 
average, less metal-enriched than the NFGS sample, the high value of $L_{\rm{[OII]}/\rm{H}\alpha}$ for the LINER/Seyfert galaxies may be attributed to 
residual star formation in these galaxies rather than the LINER/Seyfert component. 

At low redshift in cluster environments, galaxy metallicities range from $\sim$1$Z_{\odot}$ to $>5Z_{\odot}$ over the stellar mass
range of our sample [log$(M_{\star})=10^{10}-10^{11.5} M_{\odot}$; R. R. Gal et al.\ 2010, in preparation]. Because little evolution in 
the metallicity of cluster red-sequence galaxies occurs from $z\sim1$ to the present day (Kauffmann \& Charlot 1998; Kodama et al.\ 1998), the mean metallicity in
our sample is likely equal to or higher than the NFGS sample. We can test this empirically using our DEIMOS spectral data. While our spectral 
coverage does not allow us to observe H$\beta$ or [OIII] to check abundances directly through the standard R$_{23}$ diagnostic,  
we are able to constrain the \emph{F}([NeIII] $\lambda$3869\AA)/\emph{F}([OII]) ratio (hereafter $F_{\rm{[NeIII]}/\rm{[OII]}}$), which has also been shown to be 
sensitive to metal abundance (Nagao et al.\ 2006). In all galaxies in the main sample the observed $F_{\rm{[NeIII]}/\rm{[OII]}}$ or the formal 3$\sigma$ 
upper bound on this ratio is $<0.05$. Using the Nagao et al.\ (2006) relationship, this value corresponds to metallicities that are at least solar 
[log(O/H) + 12 = 8.69] and potentially much higher. While the $F_{\rm{[NeIII]}/\rm{[OII]}}$ ratio is also sensitive to the presence of an AGN, the overall 
effect of the AGN is to increase this ratio, thereby decreasing the metallicity estimate. Since our formal limit already places the NIRSPEC sample at 
equivalent or higher metallicities relative to the NFGS sample, removing the AGN contribution would simply push this limit to higher metallicities. Thus, it 
is unlikely that the higher [OII] luminosity in LINER/Seyfert galaxies is due to metallicity effects. Having ruled out dust and metallicity effects as  
causing the increased [OII] to H$\alpha$ ratio, the dominant source of the 
[OII] emission in these galaxies must come from the LINER or Seyfert component itself. Thus, \emph{the [OII] emission in such galaxies cannot be directly
tied with the star formation activity unless the LINER or Seyfert contribution is carefully subtracted}.

\subsection{Prevalence of LINER and Seyfert Activity in Cluster Galaxies}

As stated earlier, 11 out of the 19 (58\%) galaxies in the main sample have $F_{\rm{[NII]}/\rm{H}\alpha}$ ratios consistent with
at least some contribution from a LINER or Seyfert. If we include also the two ambiguous galaxies (16 and 17) that have LINER/Seyfert-like 
properties, \emph{$\sim$68\% of galaxies in the main 
sample are not consistent with pure star formation}. This fraction increases to 85\% (11/13) for red-sequence galaxies in the main sample. For bluer galaxies, 
the fraction is much less; only 33\% (2/6) of blue galaxies in the main sample have emission consistent with contributions from LINER/Seyfert sources. 

The 19 galaxies that comprise our main sample do not represent a special sub-sample of the priority 1 galaxies. 
Priority 1 targets were selected for observation only on the basis of close proximity of another priority 1 (or other high priority) object, a bias 
that is unlikely to affect their overall properties relative to the main population (see \S3.2.1). Furthermore, the objects selected in our main 
sample span a wide range of [OII] EWs and are distributed nearly over the entire color-magnitude range occupied by the whole priority 1 sample
(compare Figures \ref{fig:priorityCMD} and \ref{fig:NIRSPECCMD}). Of the 108 priority 1 targets in Cl1604, 50\% have red sequence colors. 
In RX J1821 this percentage is much higher at $\sim$87\%,
likely due to the larger fraction of red galaxies in that cluster. The difference in color properties between
the two structures is significant since the fraction of priority 1 galaxies that are LINER/Seyfert in our main sample changes as a function of color 
(i.e., 85\% of red priority 1 galaxies vs. 33\% of blue priority 1 galaxies). 

Assuming the observed fractions of LINER/Seyfert in the main sample are representative of the whole priority 1 population, 
our results suggest that \emph{$\sim$20\% of all cluster members at high-redshift with 
$M_{\star}\ga10^{10}-10^{10.5} M_{\odot}$} (our rough stellar mass completeness limit, see Appendix B) \emph{contain a LINER/Seyfert component that can be revealed by line ratios}. 
Due to the conservative nature of our LINER/Seyfert selection (i.e., our preference for purity over completeness) this number is likely a lower 
limit for galaxies in this mass range,  
increasing by as much as a factor of two if H$\beta$ and [OIII] were available in our data. Specifically, if we make a cut identical 
to the one we have used in this paper on the low-redshift data of Kauffmann et al.\ (2003b), this results in a LINER/Seyfert sample that is only 
$\sim$50\% complete. Conversely, if we were to probe less massive galaxies in the two structures
the total fraction of structure members classified as LINER/Seyfert would likely decrease. This is due to the small fractional number ($\sim10$\%) of 
galaxies with $M_{\star}<10^{10}$ at low-redshift that contain LINER or Seyfert components (Kauffmann et al.\ 2003b). 

\begin{figure*}
\plotone{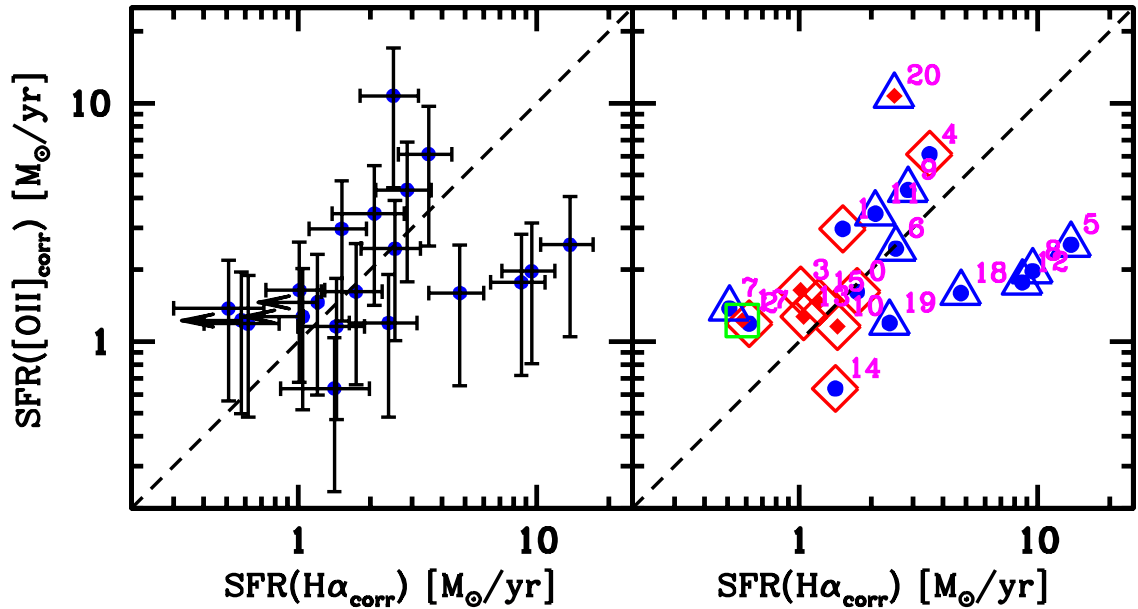}
\caption{Comparison of star formation rates (SFRs) for our Cl604 sample as derived from extinction corrected [OII] and H$\alpha$ luminosities. 
The K98 SFR conversion is used. All galaxies are corrected for internal extinction using a constant $E(B-V)=0.3$ and the Calzetti 
et al.\ (2000) reddening law. The extinction at [OII] is corrected using the extinction at the wavelength of H$\alpha$ due to the way the K98
[OII] SFR conversion was calibrated. (Left): [OII] and H$\alpha$ SFRs with error bars. Galaxies with 3$\sigma$
upper limits for L(H$\alpha$) are plotted with horizontal arrows. The dashed line marks where the SFR from the two indicators is equal. 
(Right): Similar to the left panel, but with error bars replaced by the galaxy emission class, EW properties, and galaxy numbers. 
High-[OII]/H$\alpha$ galaxies are plotted as smaller filled diamonds and low-[OII]/H$\alpha$ galaxies 
are plotted as smaller filled circles. Galaxies classified as LINER/Seyfert are plotted as large open diamonds, ambiguous galaxies are plotted as large open squares, and galaxies 
classified as star forming are plotted as large open triangles. There are two clear trends away from the unity (dashed) line. Galaxies that lie significantly to the 
right of line are all 24$\mu$m bright and are likely dust-reddened starbursts for which the effects of extinction are under-compensated. Galaxies to
the left of the line, for which the [OII] SFR is higher than the H$\alpha$ SFR, are primarily LINER/Seyfert galaxies.}
\label{fig:SFRcomp}
\end{figure*}

It is remarkable that in both Cl1604 and RX J1821, structures with 
significantly different populations, different DEIMOS selection functions, and different global spectral properties (see L09), 
the total fraction of priority 1 DEIMOS members is almost identical, 34.5\% (108/313) in Cl1604 and 31.5\% (23/73) 
in RX J1821. This fraction is, however, somewhat sensitive to the cluster sample used. If we instead cut both spectroscopic samples at our rough
completeness limit of $i\arcmin\sim23$, the fraction of priority 1 galaxies in Cl1604 increases (46.0\%) while the fraction in 
RX J1821 remains roughly constant (30.6\%). Regardless of the sample used, the fraction of priority 1 galaxies in the two systems
remains a significant fraction of the cluster galaxy population. Since a large fraction of priority 1 galaxies are likely LINER/Seyfert (i.e., $\sim$68\%), 
such galaxies seem to constitute a large fraction of the galaxy population in clusters at very different stages in their dynamical evolution. This result 
is consistent with observations of field and group galaxies at $z\sim0.8$, in which LINERs are preferentially found in denser environments (Montero-Dorta et al.\ 2009). 
The similar fraction of LINER/Seyfert galaxies in Cl1604 and RX J1821 suggests that, whatever mechanism is powering their 
emission (i.e., either a LINER or Seyfert), this mechanism is long-lived and active in cluster galaxies for much longer than the dynamical timescale of the cluster. 

Similarities in the incidence of galaxies powered by LINER/Seyfert activity can also be seen across cosmic time. At low redshift in both field and cluster environments, 
$\sim$40\% of all galaxies on the red sequence exhibit appreciable [OII] emission, 
typically due to LINER or Seyfert activity. At high redshift, in the Cl1604 supercluster environment, 54\% of all galaxies on the red sequence 
(and a similar fraction in RX J1821) have appreciable levels of [OII] emission, most of which are also likely powered by the same phase of LINER/Seyfert emission observed at low redshift. 
The similarity in the fraction of priority 1 galaxies over a large range of environments and the similarity in the fractional number of red [OII] emitters across a wide
redshift range suggests that this phenomenon is not enhanced or suppressed by the cluster environment and further reinforces the conclusion that this phase is long-lived.

\subsection{Consequences for High-redshift Galaxy Surveys}

Having established that the LINER/Seyfert population constitutes a large fraction of galaxies at both high and low redshift in both field and cluster environments, we 
now examine the consequences of our findings for large galaxy surveys at high redshift. For high redshift surveys relying on [OII] emission as a 
SFR indicator, widespread LINER/Seyfert emission can significantly bias results in several ways. The first and most obvious bias is introduced by 
incorrectly attributing [OII] emission to star formation processes rather than LINER/Seyfert emission, artificially inflating  
the measured global SFR or the star formation rate density (SFRD). A potentially more subtle bias comes when comparing the properties of 
galaxies of different spectral types in cluster or field studies as a function of local density or other environmentally sensitive parameters in order
to constrain models of galaxy evolution. In Cl1604, for example, the average EW([OII]) for our 
LINER/Seyfert sample is 18\AA, high enough to be classified as starburst or star forming in any high-redshift 
survey (Balogh et al.\ 1999; Poggianti et al.\ 2006; Franzetti et al.\ 2007; Oemler et al.\ 2009). As the spatial distribution and 
fractions of various populations (most notably K+A galaxies) are critical for many evolutionary studies in clusters, 
properly accounting for LINER/Seyferts is crucial. This is particularly important when comparing studies 
across a broad redshift range, as such quantities (e.g., SFRDs, fractional populations, etc.) derived at high redshift may be 
significantly biased relative to measurements made at low redshift where the H$\alpha$ line (or another optical recombination line) is used 
as a star formation indicator. 

To investigate the effect of the first bias on the calculated global star formation rate, we calculate the SFR of cluster 
galaxies using the relations of Kennicutt (1998, hereafter K98), due to their wide use as a conversion between the strength of nebular recombination lines and SFR.
Figure \ref{fig:SFRcomp} compares the SFR calculated using the K98 relations from the extinction corrected H$\alpha$ line and from the extinction corrected 
[OII] using a constant $E(B-V)$=0.3 (the extinction correction to [OII] is made using the extinction at H$\alpha$ due to the 
way that the K98 relations were calibrated). Two main deviations from the unity line are evident in Figure \ref{fig:SFRcomp}. The first is a population of galaxies 
at high-H$\alpha$ SFR/low-[OII] SFR, which is likely due to under-corrected extinction as most of these galaxies are 24$\mu$m bright sources. The second is a population of galaxies 
with high-[OII] SFRs relative to their H$\alpha$ SFRs, that are almost exclusively classified as LINER/Seyfert or ambiguous. 
The star formation rate as determined by the extinction corrected [OII] luminosity in the LINER/Seyfert and ambiguous galaxies 
are on average 41\% higher than those calculated using the H$\alpha$ K98 relation. This discrepancy is a 
lower limit, as three of the ten galaxies used for this comparison have H$\alpha$ SFRs calculated from 3$\sigma$ upper limits.
In contrast, the nine star-forming galaxies in this sample (excluding galaxy 20 from this analysis) have [OII] derived SFRs that are on average 8\% lower 
than SFRs calculated from H$\alpha$, consistent with no difference. 

Since both H$\alpha$ and [OII] are emitted by LINER/Seyfert-type galaxies, using either the H$\alpha$ or
[OII] SFR conversion does not accurately reflect the star formation properties of these galaxies. Though some 
contribution to the H$\alpha$ and [OII] fluxes likely come from \ion{H}{2} regions in these galaxies, especially for TOs, 
the ``contamination" of the emission lines by the LINER/Seyfert component means that a emission-line derived SFR will always be an 
overestimate of the true value. While the SFR will be overestimated regardless of whether H$\alpha$ or [OII] is used as a 
proxy, this analysis suggests that the problem becomes worse when one uses [OII] as an SFR indicator. While it has been suggested that [OII]
be abandoned as an SFR indicator in cases when H$\beta$ falls in the spectral window (Y06), corrections to the [OII] derived SFR proposed by Silverman et al.\
(2009) using the [OIII] or X-ray luminosity when available may be useful in properly accounting for the LINER/Seyfert contribution and may be 
preferable in some cases.

Finally, we analyze the effect that this population of LINER/Seyfert [OII] emitters has on the classification of 
post-starburst (i.e., K+A) galaxies. Typically, at high-redshift ($z>0.3$) a galaxy is classified as K+A based on two criteria: 
1) the presence of a strong A-star population evidenced by strong absorption in the hydrogen Balmer series (typically proxied by H$\delta$), 
and 2) no active star formation proxied by the absence of [OII] emission. These 
K+A galaxies may represent a crucial link in the transition of cluster and field galaxies from blue/late-type to red/early-type galaxies.
If we adopt one standard selection of post-starburst galaxies [e.g., EW([OII])$<5$\AA\ and EW(H$\delta)<-5$\AA; Balogh et al.\ 1999], 9.3\% (29/313) of all 
measurable Cl1604 spectra obtained with DEIMOS would be considered post-starburst, an intermediate number compared to the extreme ends of the distribution of 
K+A fractions found in other high-redshift cluster populations (Dressler et al.\ 1999; Balogh et al.\ 1999). 

While many galaxies classified as LINER/Seyfert may have some residual star formation, their optical emission profiles are dominated 
by the LINER/Seyfert component. Kauffmann et al.\ (2003b) estimated for the bright end (L$_{[OIII]}>$3.83$\times$10$^{41}$ ergs s$^{-1}$) of such 
cases, the LINER/Seyfert component contributes, on average, 50\% of the total [OII] luminosity. Thus, classification based 
on [OII] is not a sufficient criterion to rule out these galaxies as genuine post-starbursts. If we make the simple assumption that all galaxies 
classified as LINER/Seyfert have ceased forming stars and that the statistics of our main sample can be applied to the whole priority 1 population, the fraction of 
galaxies with recently truncated starburts [EW(H$\delta)<-5$\AA] in the DEIMOS Cl1604 sample increases to
18.8\% (59/313). This result suggests that 
using traditional definitions of the K+A classification \emph{severely undercounts galaxies that have recently ended their star 
formation activity, perhaps by as much as $\sim$50\%}. This value is only a rough estimate due to the over-simplified nature of the assumption 
(especially true for TOs). A more thorough investigation of the star formation histories of galaxies with LINER/Seyfert-like properties using stellar synthesis modeling 
is necessary to fully quantify the effect of H$\delta$-strong LINER/Seyfert galaxies on post-starburst selection. 

Priority 1 galaxies, of which some fraction are LINER/Seyfert ``post-starburst", have a larger range of colors (i.e., both redder 
and bluer) and absolute magnitudes than populations of ``traditional" K+A galaxies found in clusters (e.g., Poggianti et al.\ 1999; 
Dressler et al.\ 1999; see Figure \ref{fig:priorityCMD}). This disparity suggests that the LINER/Seyferts in our sample
also represent a different class of post-starburst galaxy and may have different star formation histories and progenitors than
their more traditional counterparts. The models of Poggianti et al.\ (1999), which attempted to identify the progenitors of 
traditional K+A galaxies, relied heavily on the observed magnitude and color distribution of various spectral types.
Including LINER/Seyfert post-starburst galaxies 
in such post-starburst samples would skew the overall color-magnitude distribution to redder colors and 
brighter magnitudes.

The relationship of LINER/Seyfert post-starburst galaxies to their 
traditional counterparts is not clear from these data. The fact that red priority 1 galaxies show a much higher fraction of 
LINER/Seyfert galaxies than their blue counterparts strongly suggest that these populations lie at different stages in their 
evolutionary history. The red priority 1 galaxies (of which $\sim$85\% are LINER/Seyfert) also have Balmer absorption features that 
are, on average, weaker than their blue counterparts (of which $\sim$33\% are LINER/Seyfert), suggesting that the time since the 
truncation of the star formation event is, on average, less for blue LINER/Seyfert galaxies. 
However, since we cannot discriminate between LINER/Seyfert emission and emission from a TO (the latter having ongoing star formation), 
whether or not the LINER/Seyfert mechanism is instrumental in the cessation or prevention of further star formation activity or 
whether it simply turns on after star formation has already been truncated by another process is not clear. This connection will be further investigated
in a follow-up paper using multi-wavelength data to constrain the relative ages of the stellar populations
in traditional post-starburst and LINER/Seyfert post-starburst galaxies. What is clear from our data, however, is 
that the LINER/Seyfert population in high-redshift clusters represents a substantial fraction of galaxies that are post-starburst or 
post-star-forming. Thus, in order to obtain a clear picture of galaxy evolution and to effectively link various populations in large surveys 
of galaxies it is necessary to account for contributions from LINER/Seyfert galaxies. 

\section{Conclusions}

In this study we have identified a population of [OII]-emitting, absorption-line dominated galaxies in high-redshift clusters
that are primarily powered by LINER or Seyfert activity as evidenced by their optical and NIR 
spectroscopy. Of the 486 galaxies in the Cl1604 supercluster
($z\sim0.9$) and the X-ray selected cluster RX J1821 ($z\sim0.82$) for which we have obtained optical spectroscopy, 25 galaxies were selected for follow-up 
NIRSPEC \emph{J}-band spectra. These galaxies were primarily selected to be a representative sample of a population of cluster galaxies that have optical spectra 
that exhibits moderately strong [OII] emission with no other spectral indicators of current star formation, as well as strong absorption-line features 
indicative of a well-established older stellar population. Galaxies with these spectral properties (which we have termed ``priority 1") make up a third of 
the population in both structures. The main results of this investigation are given below:

\begin{itemize}

\item We find that nearly half ($\sim$47\%) of the [OII]-emitting, absorption-line dominated galaxies in this study have high levels of [OII] emission
relative to the amount of H$\alpha$ emission. 

\item Nearly all of galaxies with high levels of [OII] emisssion relative to H$\alpha$ and a majority ($\sim$68\%) of the targeted [OII]-emitting, absorption-line 
dominated galaxies have emission profiles dominated by a LINER or Seyfert component (referred to as LINER/Seyfert), primarily revealed by the flux 
ratio of H$\alpha$ and [NII] $\lambda$6584\AA. 

\item This LINER/Seyfert fraction has a strong dependence on color; $\sim$85\% of targeted [OII]-emitting, absorption-line dominated galaxies 
on the red sequence of the two structures have dominant LINER or Seyfert components as compared with only 33\% of blue galaxies. 

\item The bulk of our LINER/Seyfert population have observed EW([OII])/EW(H$\alpha$) values significantly higher than unity, suggesting that
a majority of these galaxies are powered by LINER and not Seyfert emission. The remainder are either powered by 
Seyfert emission or undergoing a transition phase in which both LINER/Seyfert and ongoing star-forming activity is occurring. 

\item In addition to being primarily red in color, galaxies powered by LINER or Seyfert emission are almost exclusively compact early-type galaxies, contrasting 
sharply with the late-type morphologies of the star-forming galaxies observed in our sample.
 
\item The lower limit to the average 
extinction corrected \emph{L}([OII])/\emph{L}(H$\alpha$) in galaxies classified as LINER/Seyfert in our Cl1604 sample is 1.51$\pm$0.28, higher than that 
of star-forming galaxies at low redshift (i.e., 1.2; Kewley et al.\ 2004) and that of the average star-forming galaxy observed in our sample 
(1.04$\pm$0.27). We investigate various extinction schemes and metallicity differences in the samples and determine that the high levels of [OII] 
luminosity relative to H$\alpha$ in LINER/Seyfert galaxies are not due to dust or metallicity effects and are rather the result of emission from the 
LINER/Seyfert itself.

\end{itemize}

From the statistical properties of this sample we use the color distribution and prevalence of [OII]-emitting, absorption-line dominated galaxies in our 
entire Cl1604 and RX J1821 DEIMOS database
to determine the fraction of cluster galaxies at high redshift that contain a LINER/Seyfert component. For galaxies with stellar masses equal to or greater than
our stellar mass limit (roughly $M_{\star}=10^{10}-10^{10.5} M_{\odot}$) we estimate that $>20\%$ of galaxies in these structures 
contain a LINER/Seyfert component. Additionally, the fraction of galaxies on the red-sequence that have appreciable [OII] emission (most of which is likely due to 
a LINER or Seyfert) is $\sim$54\% in both structures, similar to the fraction observed in red SDSS galaxies at low redshift. Since these two systems are in significantly
different dynamical stages, these results imply that whatever mechanism is powering the emission in these galaxies is active for much longer than the dynamical 
timescale of the clusters and is not sensitive to the global environment in which a galaxy resides. 

We have established that a large fraction of high-redshift galaxies, especially those on the red sequence, have [OII] emission directly resulting
from a process unrelated to star formation. This result has significant consequences for surveys of high-redshift galaxies that use [OII] as a 
star formation indicator. The global star formation rate as calculated from the extinction corrected [OII] line luminosity for the LINER/Seyfert galaxies is significantly 
higher than the same quantity derived from the H$\alpha$ feature, itself an over-estimate of the actual SFR (due to H$\alpha$ flux originating from the 
LINER/Seyfert component). We conclude that high-redshift galaxy surveys that rely on [OII] as an SFR indicator will be non-negligibly biased by LINER/Seyfert 
activity. While other recombination lines (e.g., H$\alpha$, H$\beta$) provide better estimates of the instantaneous SFR than [OII] when observable, 
the problem of residual LINER/Seyfert flux still remains. 

We also investigate the effect of the LINER/Seyfert population on the selection of transititory ``post-starburst" galaxies, 
a population that is of considerable interest for many cluster and field evolutionary studies. We find that including H$\delta$ strong LINER/Seyfert 
galaxies increases the percentage of post-starburst galaxies in the two structures to 18.8\%, more than double the 9.3\% obtained using traditional selection methods. 
While some LINER/Seyfert galaxies likely still have some residual star formation, the requirement that [OII] be absent for a galaxy to be classified as 
post-starburst is too conservative and will result in a post-starburst sample that is severely incomplete. Due to the prevalence of LINER/Seyfert 
activity across a large range of environments at both high and low redshift, we conclude that LINER/Seyferts must be carefully accounted for when 
interpreting post-starburst populations in the context of galaxy evolution.\\[14pt]

~~ 
~~ 

We thank Jeff Newman and Michael Cooper for guidance with the \emph{spec2d} reduction
pipeline and for the many useful suggestions and modifications necessary to 
reduce our DEIMOS data. We thank Nick Konidaris for help with DEIMOS flux calibration and
for useful discussions on equivalent width measurements in DEIMOS data. We also 
thank Chris Fassnacht for providing his notes on NIRSPEC observations and the Keck II support 
astronomers for their help with the observations. B.C.L. thanks
Gary Creason for his patience and guidance. This material is based upon work supported 
by the National Aeronautics and Space Administration under Award NNG05GC34G for the Long
Term Space Astrophysics Program. Additional support for this program was provided by NASA through 
a grant HST-GO-11003 from the Space Telescope Science Institute, which is operated by the 
Association of Universities for Research in Astronomy, Inc. BCL and LML would also like to 
acknowledge support from NSF-AST-0907858. The spectrographic data presented herein were 
obtained at the W.M. Keck Observatory, which is operated
as a scientific partnership among the California Institute of
Technology, the University of California, and the National Aeronautics
and Space Administration. The Observatory was made possible by the
generous financial support of the W.M. Keck Foundation. We thank the indigenous
Hawaiian community for allowing us to be guests on their sacred mountain; we
are most fortunate to be able to conduct observations from this site.

\newpage
\begin{sidewaystable*}[tp]
\centering
\begin{center}
\caption{Cl1604 NIRSPEC Observations}
\label{tab:obs1}
\begin{tabular}{ccccccccccc}\hline \hline
$ID$ & Galaxy Number & $\alpha_{2000}$ & $\delta_{2000}$ & $z$ & \emph{F}606\emph{W} & \emph{F}814\emph{W} & \emph{F}606\emph{W}-\emph{F}814\emph{W} & Setup Number & $\tau_{\rm{exp}}$ & Class\\ \hline
J160344$+$432429 & 0  & 240.9322226 & 43.4079759 & 0.9023 &  23.60 & 21.97 & 1.63 & 1 & 5$\times$900s\tablenote{One exposure was not usable due to guider issues.} & Priority1\\ [4pt]
J160344$+$432428 & 1 & 240.9325941 & 43.4077202 & 0.9024 & 22.11 & 23.16 & 1.05 & 1 & 5$\times$900s$\tablenotemark{a}$ & Priority1\\[4pt]
J160345$+$432419 & 2  & 240.9375426 & 43.4051985 & 0.8803 &  24.01 & 22.46 & 1.55 & 1 & 5$\times$900s$\tablenotemark{a}$ & Priority1\\[4pt]
J160342$+$432406 & 3 & 240.9247136 & 43.4016956 & 0.8986 &  25.10 & 23.45 & 1.64 & 2 &  4$\times$900s & Priority1\\[4pt]
J160342$+$432403 & 4 & 240.9250684 & 43.4006981 & 0.8959 & 24.12 & 22.38 & 1.74 & 2 & 4$\times$900s & Priority1\\[4pt]
J160330$+$432208 & 5 & 240.8732075 & 43.3687725 & 0.8983 & 22.84 & 21.40 & 1.44 & 3 & 4$\times$900s & Priority1\\[4pt]
J160329$+$432204 & 6 & 240.8697693 & 43.3676967 & 0.9045 & 23.54 & 22.12 & 1.42 & 3 & 4$\times$900s & Priority1\\[4pt]
J160416$+$431021 & 7 & 241.0657080 & 43.1725670 & 0.8999 & 24.91 & 23.75 & 1.16 & 4 & 4$\times$900s & Priority3 (Filler)\\[4pt]
J160416$+$431017 & 8 & 241.0648269 & 43.1713681 &  0.8999 & 22.99 & 21.49 & 1.49 & 4 & 4$\times$900s & Priority1\\[4pt]
J160404$+$432445 & 9 & 241.0150297 & 43.4124202 & 0.9017 & 23.60 & 22.72 & 0.88 & 5 & 4$\times$900s & Priority3 (Filler) \\[4pt]
J160403$+$432436 & 10 & 241.0108301 & 43.4099384 & 0.9015 & 23.34 & 21.82 & 1.52 & 5 & 4$\times$900s & Priority1\\[4pt]
J160429$+$431956 & 11 & 241.1195420 & 43.3321920 & 0.9185 & 24.48 & 23.76 & 0.72 & 6 & 2$\times$900s & Priority3 (Filler)\\[4pt]
J160428$+$431953 & 12 & 241.1171400 & 43.3312750 & 0.9198 & 23.37 & 21.75 & 1.61 & 6 & 2$\times$900s & Priority1\\[4pt]
J160406$+$431542 & 13 &  241.0276264 & 43.2615940 & 0.8674 & 23.44 & 21.57 & 1.88 & 8 & 4$\times$900s & Priority1+Red Sequence\\[4pt]
J160407$+$431539 & 14 &  241.0299022 & 43.2607188 & 0.8676 & 24.48 & 22.81 & 1.67 & 8 & 4$\times$900s & Priority1+Red Sequence\\[4pt]
J160426$+$431423 & 15 & 241.1100259 & 43.2397136 & 0.8676 & 23.43 & 21.48 & 1.95 & 9 & 3$\times$900s & Priority1+Red Sequence\\[4pt]
J160426$+$431419 & 16 & 241.1092254 & 43.2386527 & 0.8658 & 22.80 & 20.84 & 1.96 & 9 & 3$\times$900s & Priority1+Red Sequence\\[4pt]
J160427$+$431501 & 17 & 241.1104629 & 43.2503720 & 0.8601 & 23.77 & 21.80 & 1.97 & 10 & 4$\times$900s & Priority1+Red Sequence\\[4pt]
J160426$+$431439 & 18 & 241.1086670 & 43.2441610 & 0.8710 & 24.94 & 23.35 & 1.60 & 10 & 4$\times$900s & Priority1+Red Sequence\\[4pt]
J160406$+$431825 & 19 & 241.0243330 & 43.3068500 & 0.9189 & 25.59 & 23.78 & 1.81 & 11 & 5$\times$900s$\tablenotemark{a}$ & Priority2 (Filler)\\[4pt]
J160406$+$431809 & 20 & 241.0266087 & 43.3024702 & 0.9195 & 24.29 & 22.36 & 1.93 & 11 & 5$\times$900s$\tablenotemark{a}$ & Priority1+Red Sequence\\[6pt]\hline
\end{tabular}
\end{center}
\end{sidewaystable*}

\begin{sidewaystable*}[tp]
\caption{RX J1821 NIRSPEC Observations}
\label{tab:obs2}
\begin{center}
\centering
\begin{tabular}{ccccccccccc} \hline \hline

$ID$ & Galaxy Number & $\alpha_{2000}$ & $\delta_{2000}$ & $z$ & $i\arcmin$ & $z\arcmin$ & $i\arcmin-z\arcmin$ & Setup Number & $\tau_{\rm{exp}}$ & Class\tablenote{Note that no color cut was imposed on any RX J1821 targets.} \\ \hline
J182110+682350 & 21 & 275.29224260 & 68.39710400 & 0.7960 & 21.41 & 20.76 & 0.65 & 7 & 4$\times$900s & Priority1\\[4pt]
J182108+682329 & 22 & 275.28199650 & 68.39415620 & 0.8134 & 21.75 & 21.00 & 0.75 & 7 & 4$\times$900s & Priority1\\[4pt]
J182121+682715 & 23 & 275.33619440 & 68.45408210 & 0.8092 & 23.32 & 23.03 & 0.30 & 12 & 3$\times$900s & Priority2 (filler)\\[4pt]
J182123+682714 & 24 & 275.34600740 & 68.45392090 & 0.8093 & 23.72 & 23.36 & 0.36 & 12 & 3$\times$900s & Priority2 (filler)\\[6pt]\hline
\end{tabular}
\end{center}
\end{sidewaystable*}

\begin{sidewaystable*}[tp]
\caption{Spectral Properties of the NIRSPEC Sample}
\label{tab:spec}
\begin{center}
\centering
\begin{tabular}{ccccccccc} \hline \hline

$ID$ & Galaxy Number & $z$ & EW([OII])\tablenote{Measured in the rest-frame} & EW(H$\alpha$)$\tablenotemark{a}$ & Class EW([OII])/EW(H$\alpha$)& F(H$\alpha$) & F([NII]) & Emission Class \\
 &      &               &       (\AA) &                 (\AA)               &                      & ($\times10^{-18}$ erg s$^{-1}$ cm$^{-2}$) & ($\times10^{-18}$ erg s$^{-1}$ cm$^{-2}$) & \\ \hline
J160344$+$432429 & 0 &  0.9023 &        18.8$\pm$1.7 &  14.8$\pm$2.1 &  Low &   21.8$\pm$2.9 &  17.5$\pm$2.6&                   \emph{LINER/Seyfert} \\[4pt]
J160344$+$432428 & 1 &  0.9024 &        22.5$\pm$1.3 &  13.3$\pm$1.8 &  Low &   18.9$\pm$2.1 &  13.4$\pm$2.1&                   \emph{LINER/Seyfert} \\[4pt]
J160345$+$432419 & 2 &  0.8803 &        10.7$\pm$1.3 &  6.1$\pm$1.7 &   Low  &  8.1$\pm$2.0 & 6.5$\pm$1.9&                      \emph{LINER/Seyfert} \\[4pt]
J160342$+$432406 & 3 &  0.8986 &        31.6$\pm$1.7 &  6.7$\pm$1.8 &   High &  12.7$\pm$1.6 &  17.4$\pm$2.9&                   \emph{LINER/Seyfert} \\[4pt]
J160342$+$432403 & 4 &  0.8959 &        58.6$\pm$1.4 &  17.0$\pm$1.3 &  Low &   44.5$\pm$2.2 &  53.2$\pm$3.3&                   \emph{LINER/Seyfert} \\[4pt]
J160330$+$432208 & 5 &  0.8983 &        16.1$\pm$0.8 &  45.8$\pm$1.1 &  Low &   173.0$\pm$3.6&  85.0$\pm$2.9&                   \emph{Star forming} \\[4pt]
J160329$+$432204 & 6 &  0.9045 &        8.7 $\pm$1.0 &  8.2$\pm$1.1 &   Low  &  31.4$\pm$4.1 &  13.3$\pm$2.6 &                  \emph{Star forming} \\[4pt]
J160416$+$431021 & 7 & 0.8990 &         17.4$\pm$2.3 &  38.6$\pm$14.3 & Low  &  6.4$\pm$2.1 &   $<$3.7\tablenote{3$\sigma$ upper limit} &  \emph{Star-Forming} \\[4pt]
J160416$+$431017 & 8 &  0.8999 &        17.8$\pm$1.4 &  33.3$\pm$1.7 &  Low &   119.0$\pm$4.8 & 38.4$\pm$3.1&                   \emph{Star forming} \\[4pt]
J160404$+$432445 & 9 &  0.9017 &        31.5$\pm$0.9 &  41.8$\pm$4.2 &  Low &   35.6$\pm$2.9 &  4.5$\pm$1.5&                    \emph{Star forming} \\[4pt]
J160403$+$432436 & 10 & 0.9015 &        10.2$\pm$1.9 &  3.2$\pm$0.3 &   High &  18.0$\pm$3.4 &  20.7$\pm$2.4&                   \emph{LINER/Seyfert} \\[4pt]
J160429$+$431956 & 11 & 0.9185 &        55.3$\pm$1.9 &  52.3$\pm$16.3 & Low  &  24.8$\pm$5.6 &  6.1$\pm$3.8&                    \emph{Star forming} \\[4pt]
J160428$+$431953 & 12 & 0.9198 &        9.0$\pm$1.1 &   25.4$\pm$1.8&   Low &   102.0$\pm$6.9 & 55.5$\pm$6.0&                   \emph{Star forming} \\[4pt]
J160406$+$431542 & 13 & 0.8674 &        9.9$\pm$0.9 &   0.1$\pm$0.7 &   High &  $<$14.3$\tablenotemark{b}$ &    10.6$\pm$2.4 &                  \emph{LINER/Seyfert} \\[4pt]
J160407$+$431539 & 14 & 0.8676 &        23.7$\pm$3.3 &  7.2$\pm$ 2.5  & Low &   19.4$\pm$6.1 & 12.3$\pm$4.1 &                   \emph{LINER/Seyfert} \\[4pt]
J160426$+$431423 & 15 & 0.8676 &        7.9$\pm$0.6 &   0.4$\pm$0.7  &  High &  $<$16.5$\tablenotemark{b}$  &   14.1$\pm$3.0 &                  \emph{LINER/Seyfert} \\[4pt]
J160426$+$431419 & 16 & 0.8658 &        6.3$\pm$0.7 &   1.6$\pm$1.1 &   High &  $<$22.0$\tablenotemark{b}$ & $<$27.0$\tablenotemark{b}$ &               \emph{Ambiguous} \\[4pt]
J160427$+$431501 & 17 & 0.8601 &        6.3$\pm$0.7 &   0.1$\pm$ 1.6  & High &  $<$8.1$\tablenotemark{b}$ & $<$15.7$\tablenotemark{b}$ &                        \emph{Ambiguous}  \\[4pt]
J160426$+$431439 & 18 & 0.8710 &        56.7$\pm$6.1 &  41.8$\pm$6.0 &  Low &   64.5$\pm$5.2 &  23.4$\pm$4.6 &                  \emph{Star forming} \\[4pt]
J160406$+$431825 & 19 & 0.9189 &        50.9$\pm$4.5 &  60.2$\pm$14.6 & Low &   28.4$\pm$5.5 &  13.4$\pm$5.1 &                  \emph{Star forming} \\[4pt]
J160406$+$431809 & 20 & 0.9195 &        73.7$\pm$1.4 &  14.1$\pm$3.8  & High &  29.7$\pm$3.6 &  15.8$\pm$3.9&                   \emph{Star forming} \\[4pt]
J182110$+$682350 & 21 & 0.7960 &        2.6$\pm$0.3  &  0.3$\pm$0.3 &   High &  $<$20.0$^{b}$ & 23.3$\pm$4.1 &                  \emph{LINER/Seyfert} \\[4pt]
J182108$+$682329 & 22 & 0.8134 &        26.9$\pm$0.5 &  1.9$\pm$0.5 &   High &  11.0$\pm$2.6 &  44.4$\pm$3.3 &                  \emph{LINER/Seyfert} \\[4pt]
J182121$+$682715 & 23 & 0.8092 &        17.3$\pm$1.5 &  20.6$\pm$3.5 &  Low &   37.8$\pm$6.0 &  18.1$\pm$4.8 &                  \emph{Star forming} \\[4pt]
J182123$+$682714 & 24 & 0.8093 &        24.2$\pm$2.0 &  39.9$\pm$10.0 & Low &   32.9$\pm$3.9 &  8.9$\pm$2.9 &                   \emph{Star forming} \\[6pt]\hline

\end{tabular}
\end{center}
\end{sidewaystable*}

\begin{sidewaystable*}[tp]
\caption{Star Formation Properties of the Cl1604 NIRSPEC Sample}
\label{tab:starformation}
\begin{center}
\centering
\begin{tabular}{ccccccccccc} \hline \hline

$ID$ & Galaxy & $z$ &                   F(H$\alpha$) &  F([OII]) &      L(H$\alpha$)$_{obs}$ &  L([OII])$_{obs}$ & L($H\alpha$)\tablenote{Corrected using a constant extinction of E(B-V)=0.3}$_{corr}$ & SFR(H$\alpha$)\tablenotemark{a}$_{corr}$ & L(OII)\tablenotemark{a}$_{corr}$ & SFR([OII])\tablenotemark{a}$_{corr}$ \\
 & & & & & $10^{40}$ erg s$^{-1}$ & $10^{40}$ erg s$^{-1}$ & $10^{40}$ erg s$^{-1}$ & $M_{\odot}$ yr$^{-1}$ & $10^{40}$ erg s$^{-1}$ & $M_{\odot}$ yr$^{-1}$ \\\hline
J160344$+$432429 & 0  & 0.9023 &        21.8$\pm$2.9 &  11.4 $\pm$5.1 & 8.7$\pm$1.2 &   4.6$\pm$2.1 &   22.2$\pm$6.3  &         1.7$\pm$0.5 &   23.2$\pm$11.9 &         1.6$\pm$1.0\\[4pt]
J160344$+$432428 & 1 & 0.9024 &         18.9$\pm$2.1  & 20.8$\pm$9.3  & 7.7$\pm$0.8 &   8.5$\pm$3.8 &   19.2$\pm$5.2 &          1.5$\pm$0.4 &   42.7$\pm$21.8 &         3.0$\pm$1.8 \\[4pt]
J160345$+$432419 & 2 & 0.8803  &        8.1$\pm$2.0 &   8.9$\pm$ 4.0 &  3.1$\pm$0.7 &   3.4$\pm$1.5 &   7.8$\pm$2.7&            0.6$\pm$0.2 &   17.1$\pm$8.8 &          1.2$\pm$0.7 \\[4pt]
J160342$+$432406 & 3 & 0.8986 &         12.7$\pm$1.6 &  11.6$\pm$5.2 &  5.1$\pm$0.6 &   4.7$\pm$2.1 &   12.8$\pm$3.6 &          1.0$\pm$0.3 &   23.6$\pm$12.0 &         1.6$\pm$1.0\\[4pt]
J160342$+$432403 & 4 & 0.8959 &         44.5$\pm$2.2 &  43.6 $\pm$19.3& 17.7$\pm$0.9 &  17.4$\pm$7.7 &  44.5$\pm$11.2   &       3.5$\pm$0.9 &   87.9$\pm$44.6 &         6.1$\pm$3.6 \\[4pt]
J160330$+$432208 & 5 & 0.8983 &         173.0$\pm$3.6 & 18.1$\pm$8.0 &  69.5$\pm$1.5 &  7.3$\pm$3.2 &   174.2$\pm$43.3 &        13.8$\pm$3.4 &  36.7$\pm$18.6 &         2.5$\pm$1.5 \\[4pt]
J160329$+$432204 & 6 & 0.9045 &         31.4$\pm$4.1 &  17.1$\pm$7.6 &  12.8$\pm$1.7 &  7.0$\pm$3.1 &   32.2$\pm$9.0 &          2.5$\pm$0.7 &   35.3$\pm$18.0 &         2.5$\pm$1.4 \\[4pt]
J160416$+$431021 & 7 & 0.8999 &         6.4$\pm$2.1 &   9.8 $\pm$4.3 &  2.6$\pm$0.8 &   3.9$\pm$1.7  &  6.4$\pm$2.6 &           0.5$\pm$0.2 &   19.8$\pm$10.1 &         1.4$\pm$0.8 \\[4pt]
J160416$+$431017 & 8 &  0.8999 &        119.0$\pm$4.8 & 13.9 $\pm$6.2 & 48.0$\pm$1.9 &  5.6$\pm$2.5 &   120.4 $\pm$30.2 &       9.5$\pm$2.4 &   28.4$\pm$14.5 &         2.0$\pm$1.2 \\[4pt]
J160404$+$432445 & 9 & 0.9017 &         35.6$\pm$2.9 &  30.4$\pm$13.4 & 14.4$\pm$1.2 &  12.3$\pm$5.4 &  36.2$\pm$9.4 &          2.9$\pm$0.7 &   62.1$\pm$31.5 &         4.3$\pm$2.5 \\[4pt]
J160403$+$432436 & 10 & 0.9015 &        18.0$\pm$3.4 &  8.1$\pm$3.7 &   7.3$\pm$1.4 &   3.3$\pm$1.5 &   18.3$\pm$5.7 &          1.4$\pm$0.5 &   16.6$\pm$8.5 &          1.2$\pm$0.7 \\[4pt]
J160429$+$431956 & 11 & 0.9185 &        24.8$\pm$5.6 &  23.1$\pm$10.2 & 10.5$\pm$2.4 &  9.8$\pm$4.3 &   26.3$\pm$8.8 &          2.1$\pm$0.7 &   49.5$\pm$25.1 &         3.4$\pm$2.0 \\[4pt]
J160428$+$431953 & 12 & 0.9198 &        102.0$\pm$6.9 & 11.8$\pm$5.3 &  43.4$\pm$2.9 &  5.0$\pm$2.3 &   108.9$\pm$ 27.9 &       8.6$\pm$2.2 &   25.4$\pm$13.1 &         1.8$\pm$1.1 \\[4pt]
J160406$+$431542 & 13  & 0.8674 &       $<$14.3\tablenote{3$\sigma$ upper limit} & 9.8$\pm$4.4 &  $<$5.3$\tablenotemark{b}$ &     3.6$\pm$1.6 &   $<$13.2$\tablenotemark{b}$ &                 $<$1.0$\tablenotemark{b}$ &  18.3$\pm$9.4 &          1.3$\pm$0.8 \\[4pt]
J160407$+$431539 & 14  & 0.8676 &       19.4$\pm$6.1 &  4.9$\pm$2.4 &   7.2$\pm$2.3 &   1.8$\pm$0.9 &   17.9$\pm$7.2 &          1.4$\pm$0.6 &   9.1$\pm$5.1 &           0.6$\pm$0.4 \\[4pt]
J160426$+$431423 & 15 & 0.8676 &        $<$16.5$\tablenotemark{b}$ &    11.3$\pm$5.0 &  $<$6.1\tablenotemark{b} &     4.2$\pm$1.9 &   $<$15.2\tablenotemark{b} &   $<$1.2$^{b}$ &  21.0$\pm$10.7 &         1.5$\pm$0.9 \\[4pt]
J160426$+$431419 & 16 & 0.8658 &        $<$22.0\tablenotemark{b} &      ...\tablenote{No flux measurement was attempted for the LRIS object, due to the uncertainty in the flux calibration.} &  $<$8.1\tablenotemark{b} & ...\tablenotemark{c} & $<$20.2\tablenotemark{b} & $<$1.6\tablenotemark{b} & ...\tablenotemark{c} & ...\tablenotemark{c} \\[4pt]
J160427$+$431501 & 17 & 0.8601 &        $<$8.1\tablenotemark{b} &       9.7$\pm$4.4 &   $<$2.9\tablenotemark{b} &       3.5$\pm$1.6 &   $<$7.3\tablenotemark{b} &               $<$0.6\tablenotemark{b} &       17.7$\pm$9.1 &          1.2$\pm$0.8 \\[4pt]
J160426$+$431439 & 18 & 0.8710 &        64.5$\pm$5.2 &  12.2$\pm$5.5 &  24.0$\pm$1.9 &  4.6$\pm$2.0 &   60.2$\pm$15.7 &         4.8$\pm$1.2 &   23.0$\pm$11.8 &         1.6$\pm$0.9 \\[4pt]
J160406$+$431825 & 19 & 0.9189 &        28.4$\pm$5.5 &  8.0$\pm$3.6 &   12.0$\pm$2.4 &  3.4$\pm$1.5 &   30.3$\pm$9.5 &          2.4$\pm$0.8 &   17.2$\pm$8.9 &          1.2$\pm$0.7 \\[4pt]
J160406$+$431809 & 20 & 0.9195 &        29.7$\pm$3.6 &  7.2$\pm$3.2  &  12.6$\pm$1.5 &  30.6$\pm$13.5 & 31.7$\pm$8.7    &       2.5$\pm$0.7&    154.4$\pm$78.3 &        10.7$\pm$6.3 \\[6pt]\hline

\end{tabular}
\end{center}
\end{sidewaystable*}

\begin{sidewaystable*}[tp]
\caption{Imaging properties of the NIRSPEC sample}
\label{tab:imaging}
\begin{center}
\centering
\begin{tabular}{ccccccccccccccccc} \hline \hline

$ID$ & Galaxy Number & $z$ & $\alpha_{2000}$ & $\delta_{2000}$ &  $m_{F606W}$ & $m_{F814W}$ & r$\arcmin$ & i$\arcmin$ & z$\arcmin$ & K$_s$ & Color\tablenote{As defined in \S5.3} & Morphology\tablenote{Done by visual inspection, M = Merger, I = Interaction, C = Chaotic, S = Spiral, Asymm = Asymmetric Disk}\\ \hline

J160344$+$432429 & 0 & 0.9023 & 240.9322226 & 43.4079759  & 23.6011 & 21.9680 & 23.0289 & 21.6730 & 20.9049 & ...\tablenote{Not detected in K$_{s}$} & Red & S M I\\[4pt]
J160344$+$432428 & 1 & 0.9024 & 240.9325941 & 43.4077202 & 23.7290 & 22.1054 & 23.1568 & 21.8849 & 21.2576 & 20.5956 & Red & S C I\\[4pt]
J160345$+$432419 & 2 & 0.8803 & 240.9375426 & 43.4051985 & 24.0052 & 22.4560 & 23.7045 & 22.6779 & 22.0024 & 20.7284 & Blue & S0 \\[4pt]
J160342$+$432406 & 3 & 0.8986 & 240.9247136 & 43.4016956 & 25.0951 & 23.4488 & 23.9627 & 22.7284 & 22.2418 & 21.1584 & Blue & E \\[4pt]
J160342$+$432403 & 4 & 0.8959 & 240.9250684 & 43.4006981 & 24.1215 & 22.3766 & 23.4728 & 22.3332 & 21.5753 & 20.4692 & Red & E elongated \\[4pt]
J160330$+$432208 & 5 & 0.8983 & 240.8732075 & 43.3687725 & 22.8355 & 21.3960 & 22.4344 & 21.5230 & 20.8415 & 19.4092 & Blue & Sc I\\[4pt]
J160329$+$432204 & 6 & 0.9045 & 240.8697693 & 43.3676967 & 23.5419 & 22.1219 & 22.9297 & 21.8290 & 20.9793 & 20.5688 & Blue & Sa Ring\\[4pt]
J160416$+$431021 & 7 & 0.8990 & 241.0657080 & 43.1725670 & 24.9100 & 23.7452 & 24.9532 & 24.0943 & 22.4512 & ...$^{\rm{c}}$ & Blue& Amorphous \\[4pt]
J160416$+$431017 & 8 & 0.8999 & 241.0648269 & 43.1713681 & 22.9863 & 21.4923 & 22.4896 & 21.2475 & 20.4742 & 19.1758 & Blue & Sc I \\[4pt]
J160404$+$432445 & 9 & 0.9017 & 241.0150297 & 43.4124202 & 23.5993 & 22.7161 & 23.8893 & 22.8681 & 22.7997 & ...$^{\rm{c}}$ & Blue & S asymm M?\\[4pt]
J160403$+$432436 & 10 & 0.9015 & 241.0108301 & 43.4099384 & 23.3368 & 21.8190 & 23.0297 & 21.9360 & 21.2114 & 19.7736 & Blue & Sa/S0 \\[4pt]
J160429$+$431956  & 11 & 0.9185 & 241.1195420 & 43.3321920  & 24.4849 & 23.7607 & 25.1182 & 23.8742 & 23.8430 & ...$^{\rm{c}}$ & Blue & S0 peculiar I? \\[4pt]
J160428$+$431953 & 12 & 0.9198 & 241.1171400 & 43.3312750 & 23.3654 & 21.7545 & 23.1085 & 21.9486 & 21.2242 & 19.5014 & Blue & SBb \\[4pt]
J160406$+$431542 & 13 & 0.8674 & 241.0276264 & 43.2615940 & 23.4426 & 21.5654 & 22.5928 & 21.5389 & 20.7046 & 19.2990 & Red & S0 asymm\\[4pt]
J160407$+$431539 & 14 & 0.8676 & 241.0299022 & 43.2607188 & 24.4806 & 22.8149 & 23.9034 & 22.9086 & 22.1914 & 20.9843 & Red & elongated E \\[4pt]
J160426$+$431423 & 15 & 0.8676 & 241.1100259 & 43.2397136 & 23.4286 & 21.4794 & 21.9561 & 20.9147 & 19.9575 & 18.7911 & Red & E I? \\[4pt]
J160426$+$431419 & 16 & 0.8658 & 241.1092254 & 43.2386527  & 22.7983 & 20.8419 & 21.6547 & 20.3484 & 19.4208 & 18.3566 & Red & E \\[4pt]
J160427$+$431501 & 17 & 0.8601 & 241.1104629 & 43.2503720 & 23.7716 & 21.8000 & 22.8843 & 21.5115 & 20.5948 & 19.2840 & Red & E I? \\[4pt]
J160426$+$431439 & 18 & 0.8710 & 241.1086670 & 43.2441610 & 24.9447 & 23.3475 & 25.0073 & 23.4083 & 22.1090 & 21.3286 & Red & Amorphous?\\[4pt]
J160406$+$431825 & 19 & 0.9189 & 241.0243330 & 43.3068500 & 25.5927 & 23.7835 & 25.4611 & 23.9588 & 22.9134 & 21.2724 & Red & S disturbed \\[4pt]
J160406$+$431809 & 20 & 0.9195 & 241.0266087 & 43.3024702  & 24.2884 & 22.3585 & 22.7956 & 22.0227 & 21.1015 & 19.8396 & Red & E \\[4pt]
J182110+682350 & 21 & 0.7960 & 275.2922426 & 68.3971040 & ...\tablenote{ACS data not available for RX J1821} & ...$^{\rm{d}}$ & 22.5616 & 21.4127 & 20.7620 & ...\tablenote{$K_{s}$ magnitudes not available for RX J1821} & Red & ...$^{\rm{d}}$ \\[4pt]
J182108+682329 & 22 & 0.8134 & 275.2819965 & 68.3941562 & ...$^{\rm{d}}$ & ...$^{\rm{d}}$ & 22.8643 & 21.7501 & 20.9960 & ...$^{\rm{e}}$ & Red & ...$^{\rm{d}}$ \\[4pt]
J182121+682715 & 23 & 0.8092 & 275.3361944 & 68.4540821 & ...$^{\rm{d}}$ & ...$^{\rm{d}}$ & 24.1828 & 23.3243 & 23.0264 & ...$^{\rm{e}}$ & Blue & ...$^{\rm{d}}$ \\[4pt]
J182123+682714 & 24 & 0.8093 & 275.3460074 & 68.4539209 & ...$^{\rm{d}}$ & ...$^{\rm{d}}$ & 24.4237 & 23.7167 & 23.3609 & ...$^{\rm{e}}$ & Red & ...$^{\rm{d}}$ \\[6pt]\hline

\end{tabular}
\end{center}
\end{sidewaystable*}

\begin{figure*}[p]
\plotone{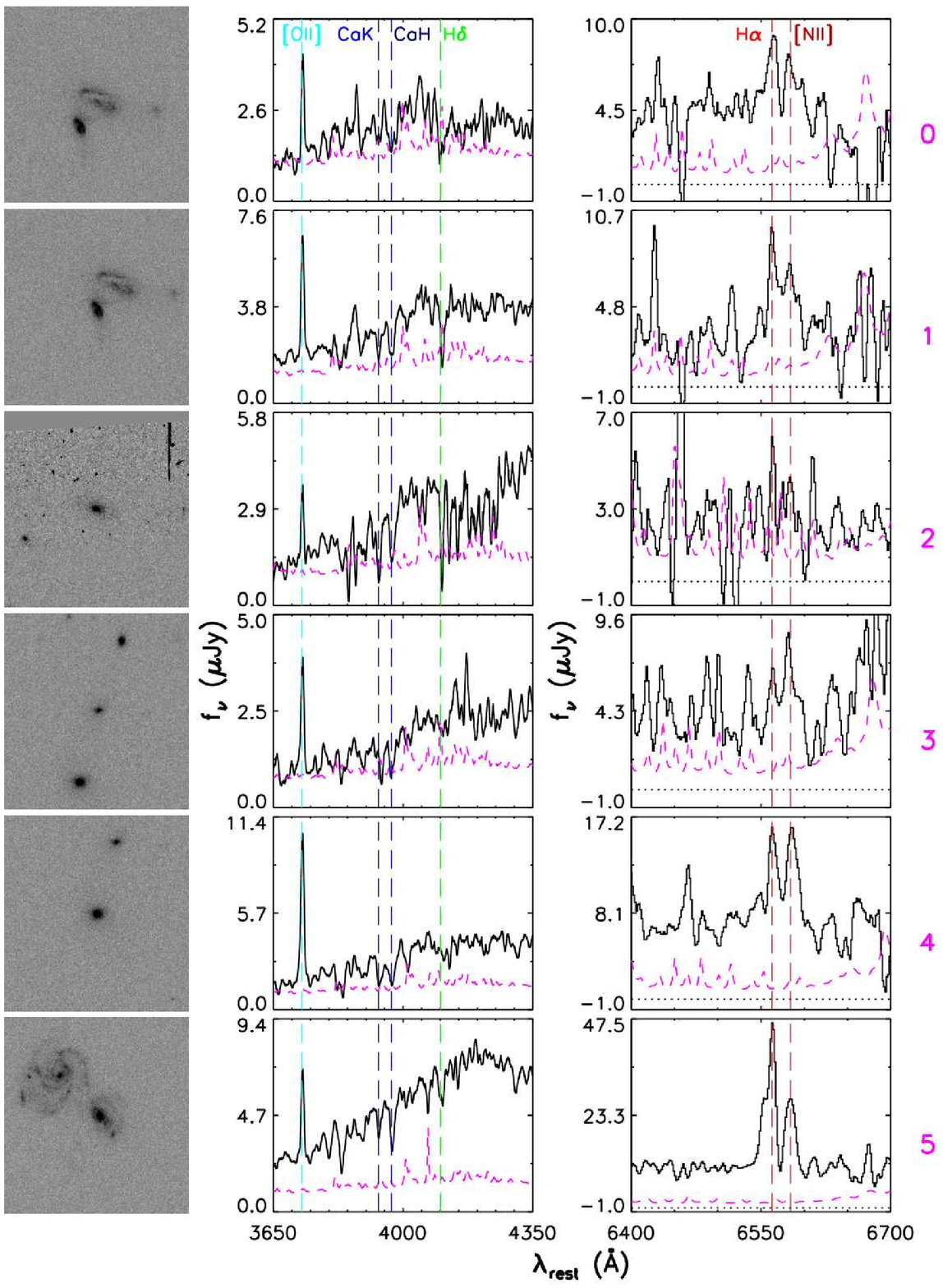}
\caption{ACS \emph{F}814\emph{W} postage stamps of each galaxy targeted with NIRSPEC as well as the associated rest-frame DEIMOS (left) 
and NIRSPEC (right) spectra of each galaxy. The galaxy number is indicated next to each spectrum. The DEIMOS 
spectra are smoothed with a Gaussian of FWHM 15 pixels (roughly 2.7 \AA\ rest-frame at the redshift of the supercluster) 
and the NIRSPEC spectra are smoothed with a Gaussian of FWHM 1.7 pixels (roughly 2.7 \AA\ rest-frame). The error spectrum
is plotted with a dashed line below each DEIMOS and NIRSPEC spectrum. All spectra are flux calibrated; however, no correction is made 
for internal extinction. For clarity the DEIMOS spectra are plotted with the zero flux level at the bottom of the plot. 
Due to the low level of continuum emission in some of the NIRSPEC targets, a dotted line shows the zero flux level for each NIRSPEC
spectrum. The long dashed lines show the locations of important spectral features. NIRSPEC targets 0-5.}
\label{fig:mosaic1}
\end{figure*}

\begin{figure*}[p]
\plotone{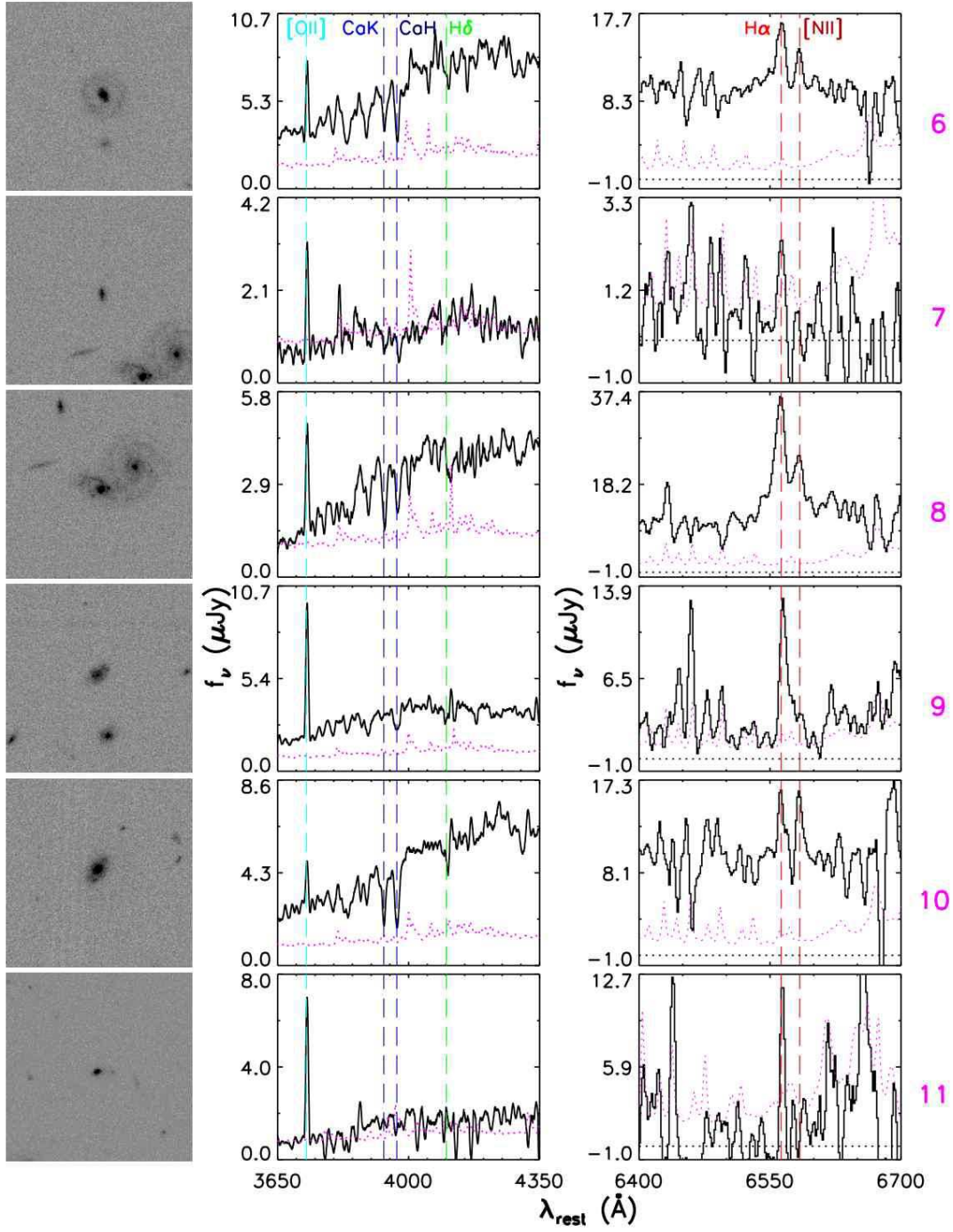}
\caption{Same as in Figure \ref{fig:mosaic1}. NIRSPEC targets 6-11.}
\label{fig:mosaic2}
\end{figure*}

\begin{figure*}[p]
\plotone{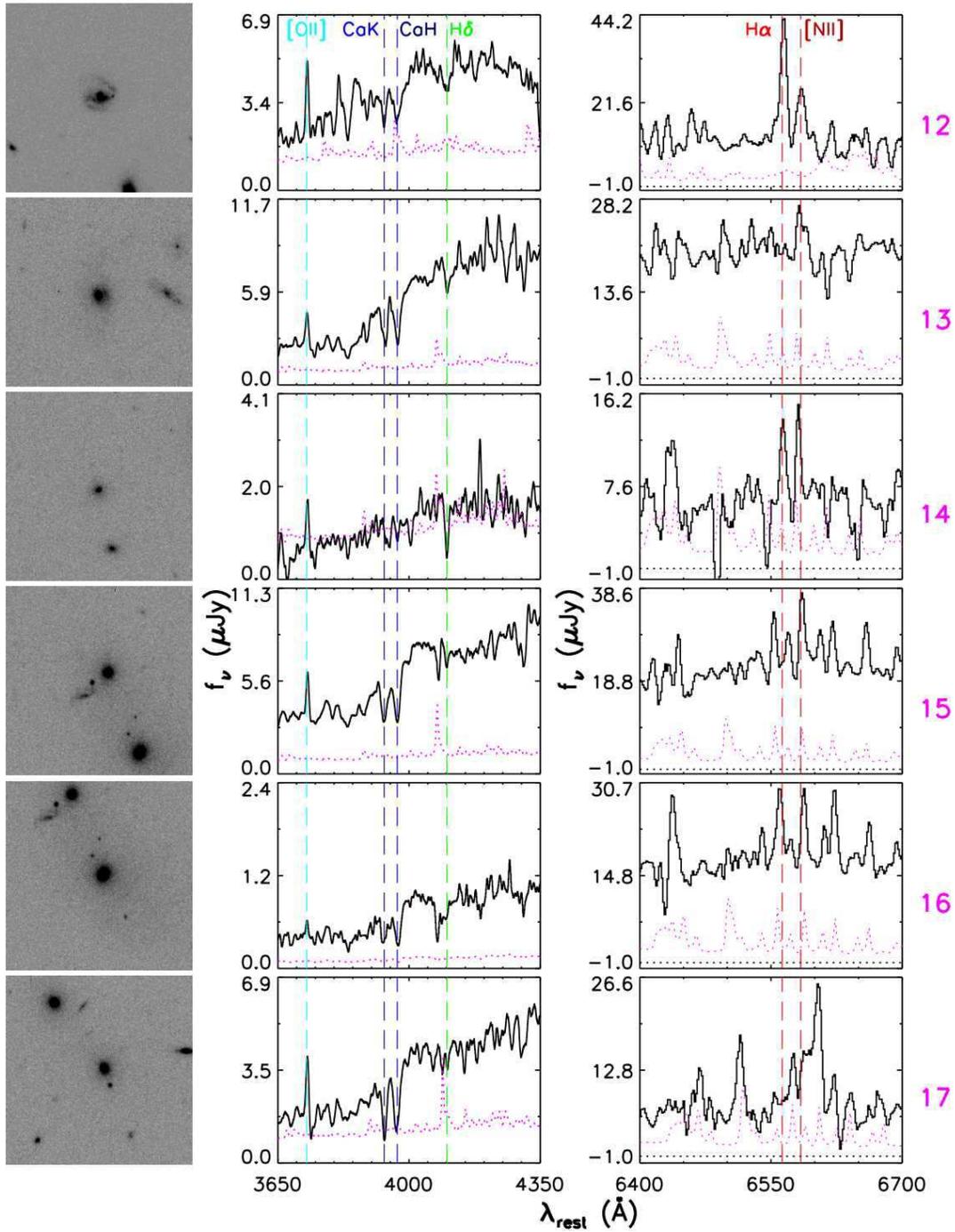}
\caption{Same as in Figure \ref{fig:mosaic1}. NIRSPEC targets 12-17. The spectrum for galaxy 16 obtained with LRIS (center panel) is not flux calibrated. The spectrum is smoothed with a Gaussian with a 2 pixel FWHM (roughly 3.4 \AA\ rest-frame).}

\label{fig:mosaic3}
\end{figure*}

\begin{figure*}[p]
\plotone{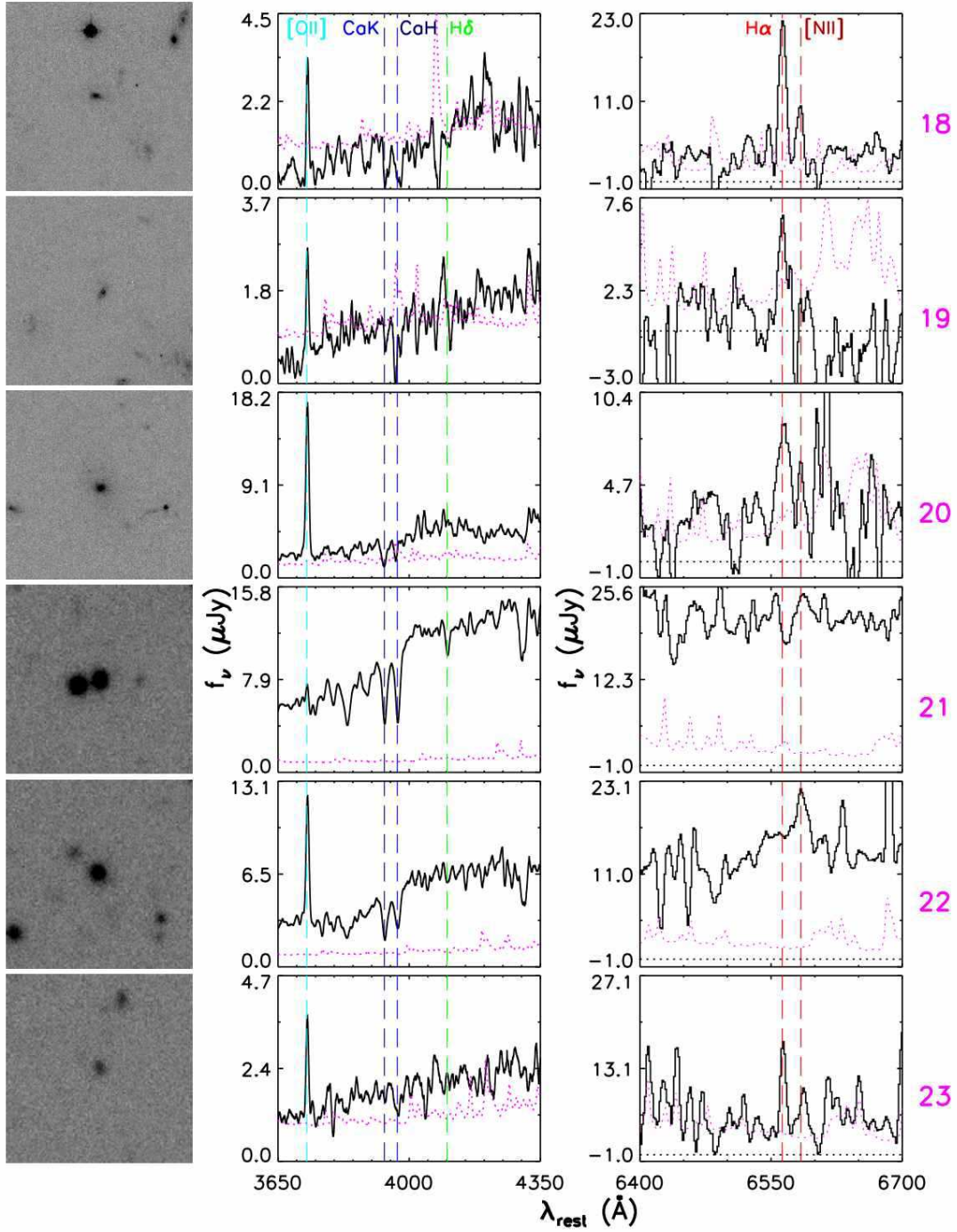}
\caption{Same as in Figure \ref{fig:mosaic1}. NIRSPEC targets 18-23. LFC $i\arcmin$ postage stamps are used for galaxies 21-23 due to the lack of ACS data in the RX J1821 field.}
\label{fig:mosaic4}
\end{figure*}

\begin{figure*}[p]
\plotone{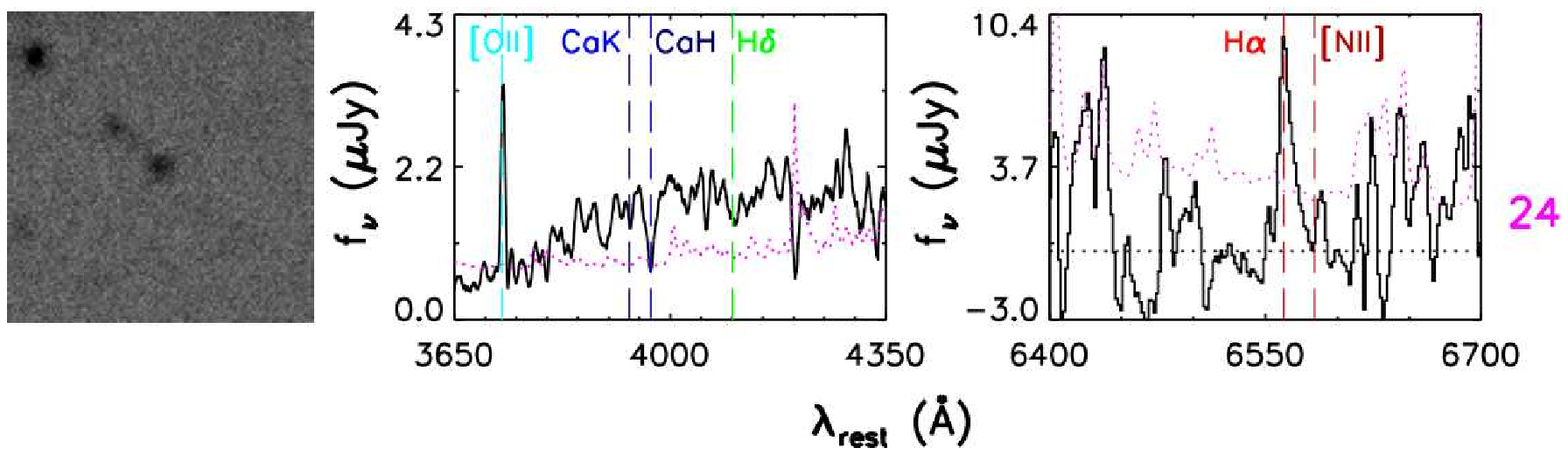}
\caption{Same as in Figure \ref{fig:mosaic1}. NIRSPEC target 24. A LFC $i\arcmin$ postage stamp is used for galaxy 24 due to the lack of ACS data in the RX J1821 field.}
\label{fig:mosaic5}
\end{figure*}

\appendix
\section{\normalsize{\bf{Appendix A:} Equivalent Width Measurements}}

In order to measure the rest-frame EW, the wavelength values of each observed-frame spectrum are divided by (1+$z$).
In all cases the DEIMOS/LRIS redshift is used. In most cases there were no significant differences between the
NIRSPEC and DEIMOS redshifts, and the maximal offsets of $\Delta z = 0.0005-0.002$ ($\Delta$v$\approx100-300$ km s$^{-1}$)
have little effect on the EW measurement.

Line-fitting EW measurements were performed for all DEIMOS spectra, as our entire NIRSPEC sample had [OII] lines 
detected at a significance of greater than
3$\sigma$. For these spectra, we fit a double Gaussian model plus a linear continuum to the 3726\AA\ and 3729\AA\ [OII] doublet,
which is typically resolved by the DEIMOS 1200 l mm$^{-1}$ grating. In cases where the [OII] doublet was not resolved
a double Gaussian model was still used. As these measurements were done in the rest frame, the two Gaussians were fixed to a
separation of 2.8\AA. Thus, the model contained seven free parameters, two to characterize the linear continuum,
four to characterize the FWHM and amplitude of each Gaussian, and a single parameter defining the mean wavelength of the
blueward Gaussian.

For NIRSPEC data, we fit only those spectra where both H$\alpha$ and [NII] were detected
at greater than 3$\sigma$. Again, we use a seven parameter double Gaussian plus linear continuum
to fit H$\alpha$ and [NII], with a fixed separation of 20.6\AA. As noted in the text (see \S4.1), adding a third Gaussian to account for the 
blueward [NII] $\lambda$6548\AA\ feature had a negligible effect on the EW measurements. In all cases where fitting was used to determine
EWs, errors were estimated from the covariance matrix of the fit. 

Bandpass measurements were performed by defining two ``continuum" bandpasses, slightly blueward and redward of the spectral feature, 
which are used to estimate the stellar continuum across the emission feature. An additional ``feature" bandpass is defined to
encompass the spectral line. A $\chi^{2}$ minimization to the linear continuum terms was performed over the
two continuum bandpasses. Any pixels with large variance values (typically from bright sky features) were removed from the continuum
bandpasses. We do not remove similar pixels in the feature bandpass. The EW is defined as:

\begin{equation}
EW [\rm{\AA}] = \sum_{i=0}^n \frac{F_{i} - C_{i}}{C_{i}}\Delta\lambda_{r,i}
\label{eqn:EW}
\end{equation}

\noindent where F$_{i}$ is the flux in the $i$th pixel in the feature bandpass, $C_{i}$ is the continuum flux in the $i$th pixel over the same
bandpass, and $\Delta\lambda_{r,i}$ is the restframe pixel scale of the spectrum (in \AA\ pixel$^{-1}$). Errors in the EW were derived using a
combination of Poisson errors on the spectral feature and the covariance matrix of the linear continuum fit and are given by (Bohlin et al.\ 1983):

\begin{equation}
\sigma_{EW} [\rm{\AA}] = \sqrt{\left(\sum_{i=0}^n \frac{\sigma_{F,i}\Delta\lambda_{i} }{C_{i}}\right)^{2} + \left(\sigma_{C}\sum_{i=0}^n\frac{F_{i}\Delta\lambda_{i}}{C_{i}^2}\right)^2},
\label{eqn:EWerr}
\end{equation}

Bandpasses were initially chosen to be ``standard", using the bandpasses of Fisher et al.\ (1998) for the [OII] feature (blue continuum:
[3696.3, 3716.3], red continuum: [3738.3, 3758.3], feature bandpass: [3716.3, 3738.3]) and the bandpasses of Y06 for the H$\alpha$ and
[NII] features (for both features, blue continuum: [6483.0, 6513.0], red continuum: [6623.0,6653]; H$\alpha$ feature bandpass: [6554.6, 6574.6], [NII]
feature bandpass: [6575.3, 6595.3]). These bandpasses were then modified by eye for each galaxy spectrum to avoid poorly
subtracted airglow lines and to avoid ``contaminate" features near the spectral lines of interest (primarily higher order
Balmer lines when measuring [OII]).

For high S/N spectral features, line-fitting techniques generally gave more accurate values and smaller errors for the EW, as noise
in the data has a relatively small effect on the overall fit (Goto et al.\ 2003; Y06; Tremonti et al.\
2004). For lower S/N lines bandpass measurements generally led to more accurate
results (Goto et al.\ 2003; Y06). Many of the spectral features that we are measuring (especially in the
NIRSPEC data) have pixel S/N $\la$5, and these data can be severely affected by over- or under-subtracted
skylines. Therefore, we chose the EW measurements derived from bandpass techniques for a majority of EWs.
In all cases where measurements of the [OII], H$\alpha$, and [NII] features
were made using both techniques the two methods agreed within the errors for high S/N features. Only at low S/N, when the
line-fitting technique began to fail, did the EW measurement differ appreciably between the two methods.

\section{\normalsize{\bf{Appendix B:} Extinction Correction Methods and Stellar Masses}}

The problem of extinction correction in this data set is complicated by the nature of our sample. Many conventional correction
methods, such as those mentioned in section \S4.2.3, are made assuming the dominant contribution to the recombination lines comes from
\ion{H}{2} regions rather than LINERs or Seyferts. As many of the galaxies in our sample contain either dominant LINER/Seyfert emission 
or a linear combination of LINER/Seyfert and star formation activity, assumptions such as a mean Balmer decrement or average 
observed [OII]/H$\alpha$ ratios are not necessarily valid for our sample. While we adopt a constant extinction value of $E(B-V)$=0.3 for 
our data (see \S4.2.3), we report here on the three methods that were used to constrain our choice of $E(B-V)$=0.3 and to justify its use.

Extinction estimates from the 24$\mu$m data were made by comparing the SFR
calculated through a linear combination of the observed H$\alpha$ and 24$\mu$m luminosities using the formula of Calzetti
et al.\ (2007). This value is compared with the value of the SFR measured using H$\alpha$ alone. \emph{K}-corrections
from the observed MIPS luminosity to the rest-frame 24$\mu$m were derived using the templates of Chary \& Elbaz (2001).
For the six galaxies with 24$\mu$m detections in the Cl1604 NIRSPEC sample, the $E(B-V)$ values range from $E(B-V)$=0.15
to 0.71, with a mean of 0.32$\pm$0.09.

The second method is based on absolute \emph{B}-band magnitudes, which are estimated
by \emph{K}-correcting the observed i$\arcmin$ magnitude (nearly identical to the Johnson \emph{B} band at the supercluster
redshift) of each Cl1604 NIRSPEC target. Extinction values were generated using the
Argence \& Lamareille (2009) adaptation of the best-fit $M_{B}-E(B-V)$ relationship of Moustakas et al.\ (2006). The
derived $E(B-V)$ values range from 0 to 0.66, with a mean of 0.29$\pm$0.07 for the 20 galaxies in the
Cl1604 NIRSPEC sample for which \emph{K}-corrections could be performed.

The third estimate of extinction was derived from synthetic stellar template fits to the optical/IR SED using the
Le PHARE\footnote{http://www.oamp.fr/people/arnouts/LE\_PHARE.html} (Arnouts \& Ilbert) codes with the single-burst stellar population
models of Bruzual \& Charlot (2003).
Using the redshift as a prior, the Le PHARE code provides an estimate of the stellar mass, stellar age, extinction,
metallicity, and $\tau$ (the e-folding time of a single star formation event) for each galaxy. The stellar mass
of Cl1604 members range from $M_{\star}=10^{9} M_{\odot}$ to $10^{11.5} M_{\odot}$ (the NIRSPEC targets range from $M_{\star}=10^{10}-10^{11.5} M_{\odot}$). Our completeness limit, roughly proxied by the turnover in the supercluster galaxy mass function, corresponds to $M_{\star}=10^{10}-10^{10.5} M_{\odot}$.
Extinction values were only used for galaxies that were cleanly detected in $r\arcmin i\arcmin z\arcmin 
K_{s}$ and at least the first two IRAC channels (3.6 and 4.5 $\mu$m). This criterion is necessary as strong degeneracies exist between extinction and the other estimated parameters
(e.g. metallicity and age) that are difficult to break without detections in the first two IRAC channels. The extinction values estimated for
the 17 Cl1604 NIRSPEC targets that were detected in all six bands range from $E(B-V)$ = 0 to 0.4, with a mean of 0.24$\pm$0.02.

\end{document}